\documentclass[useAMS,usenatbib]{mnras}
\usepackage{aas_macros}
\usepackage{graphicx}
\usepackage{amsmath,amssymb}
\usepackage[T1]{fontenc}
\usepackage{aecompl}
\usepackage{times}
\title[Baryons in the Cosmic Web -- I]{Baryons in the Cosmic Web of IllustrisTNG -- I: Gas in Knots, Filaments, Sheets and Voids}
\author[D. Martizzi et al.]{\parbox[t]{\textwidth}{Davide Martizzi$^{1,2}$\thanks{E-mail: davide.martizzi@nbi.ku.dk}, Mark Vogelsberger$^{3}$, Maria Celeste Artale$^{4}$, Markus Haider$^{4}$, Paul Torrey$^{3}$, Federico Marinacci$^{3}$, Dylan Nelson$^{5}$, Annalisa Pillepich$^{6}$, Rainer Weinberger$^{7}$, Lars Hernquist$^{8}$, Jill Naiman$^{8}$, Volker Springel$^{6,7,9}$} \\ \\
$^{1}$DARK, Niels Bohr Institute, University of Copenhagen, 2100 Copenhagen, Denmark \\
$^{2}$Department of Astronomy and Astrophysics, University of California, Santa Cruz, CA 95064, USA \\
$^{3}$MIT Kavli Institute for Astrophysics \& Space Research, Cambridge, MA, 02139, USA \\
$^{4}$Institut f\"{u}r Astro- und Teilchenphysik, Universit\"{a}t Innsbruck, Technikerstrasse 25/8, 6020 Innsbruck, Austria \\
$^{5}$Max-Planck-Institut f\"{u}r Astrophysik, Karl-Schwarzschild-Str. 1, 85741 Garching, Germany \\
$^{6}$Max-Planck-Institut f\"{u}r Astronomie, K\"{o}nigstuhl 17, 69117 Heidelberg, Germany \\
$^{7}$Heidelberg Institute for Theoretical Studies, Schloss-Wolfsbrunnenweg 35, D-69118 Heidelberg, Germany \\
$^{8}$Harvard-Smithsonian Center for Astrophysics, 60 Garden Street, Cambridge, MA 02138 \\
$^{9}$Zentrum f\"{u}r Astronomie der Universit\"{a}t Heidelberg, Astronomisches Recheninstitut, M\"{o}nchhofstr. 12-14, 69120, Heidelberg, Germany \\
}

\begin{document}

\maketitle

\begin{abstract}
We analyze the IllustrisTNG simulations to study the mass, volume fraction and phase distribution of gaseous baryons embedded in the knots, filaments, sheets and voids of the Cosmic Web from redshift $z=8$ to redshift $z=0$. We find that filaments host more star-forming gas than knots, and that filaments also have a higher relative mass fraction of gas in this phase than knots. We also show that the cool, diffuse Intergalactic Medium (IGM; $T<10^5 \, {\rm K}$, $ n_{\rm H}<10^{-4}(1+z) \, {\rm cm^{-3}}$) and the Warm-Hot Intergalactic Medium (WHIM; $ 10^5 \, {\rm K} <T<10^7 \, {\rm K}$, $ n_{\rm H} <10^{-4}(1+z)\, {\rm cm^{-3}}$) constitute $\sim 39\%$ and $\sim 46\%$ of the baryons at redshift $z=0$, respectively. Our results indicate that the WHIM may constitute the largest reservoir of {\it missing} baryons at redshift $z=0$. Using our Cosmic Web classification, we predict the WHIM to be the dominant baryon mass contribution in filaments and knots at redshift $z=0$, but not in sheets and voids where the cool, diffuse IGM dominates. We also characterise the evolution of WHIM and IGM from redshift $z=4$ to redshift $z=0$, and find that the mass fraction of WHIM in filaments and knots evolves only by a factor $\sim 2$ from redshift $z=0$ to $z=1$, but declines faster at higher redshift. The WHIM only occupies $4-11\%$ of the volume at redshift $0\leq z \leq 1$. We predict the existence of a significant number of currently undetected OVII and NeIX absorption systems in cosmic filaments which could be detected by future X-ray telescopes like Athena. 
\end{abstract}


\section{Introduction} \label{sec:intro}

The large scale structure of the observable Universe is organised in a web-like structure that is often referred to as Cosmic Web. The basic properties of the Cosmic Web can be predicted by relatively simple theories like the Zel'dovich approximation \citep{2014MNRAS.439.3630W}, whereas many aspects of its structure are predicted in exquisite detail by cosmological N-body simulations that follow the evolution of dark matter dynamics in cosmological volumes \citep{1996Natur.380..603B, 1999ApJ...526..568S, 2003ApJ...599L...1F, 2005ApJ...632...58R}. 

The Cosmic Web is organised in a hierarchy of inter-connected structures formed via gravitational instability in the expanding Universe. Most of the observable objects in the Universe reside in structures that gravitationally collapsed along 3 principal axes, the knots of the Cosmic Web, forming dark matter halos. Halos are connected to each other by filaments, which are gravitationally collapsed only along 2 principal axes. Large sheets of matter can also form, when a region of the Universe collapses only along one principal direction. The volume of the Universe between knots, filaments and sheets is filled by cosmic voids, large under-dense patches. The structure of the Cosmic Web, the volume and mass fractions belonging to each component (knots, filaments, sheets, voids) evolve dramatically from high to low redshift \citep{2007MNRAS.375..489H, 2009MNRAS.393..457S, 2014MNRAS.441.2923C}. 

At redshift $z<0.5$ the Cosmic Web predicted by numerical simulations and reconstructed from galaxy redshift surveys exhibits very extended voids and large mass fractions in knots and filaments. Although this structure is largely determined by the dynamics of dark matter, it is quite important to characterise the state of baryonic matter and the way it is distributed in the Cosmic Web, because this is the only component that can be directly observed. 

A census of baryonic matter in all observable phases has been initially compiled by \cite{1998ApJ...503..518F} and revised in the following years \citep[][]{2004IAUS..220..227F, 2007ARA&A..45..221B}. The baryonic matter abundance can be constrained by measurements of the Lyman$-\alpha$ forest at redshift $z>1$ \citep{1997ApJ...489....7R, 2003ApJS..149....1K}, which traces most of the cosmological baryons in the Universe at those epochs and which has been recently used to perform tomography of the Cosmic Web \citep[e.g.][]{2014ApJ...795L..12L, 2018ApJS..237...31L}. The baryonic matter abundance and spatial distribution is well constrained by multi-wavelength observations of galaxies and their proximity at redshift $z<5$, in particular in galaxy groups and clusters \citep[][]{2004ApJ...617..879L, 2009ApJ...703..982G, 2012ApJ...744..159L, 2013ApJ...778...14G, 2013A&A...551A..23E}, and in the Circumgalactic Medium of galaxies (CGM; HI absorption lines quasar spectra; \citealt{2012ApJ...750...67R, 2017ApJ...837..169P}; CII and CIV absorption lines in quasar spectra; \citealt{2014ApJ...796..140P, 2017ApJ...846..151H, 2018arXiv180909777S}; OVI absorption lines in quasar spectra; \citealt{2006ApJ...637..648S, 2014ApJ...792....8W, 2015MNRAS.449.3263J}; MgII absorption lines in quasar spectra; \citealt{2008AJ....135..922K, 2012ApJ...760L...7K,2013ApJ...776..115N,2013ApJ...763L..42C, 2018ApJ...853...95R}; SiIII absorbers in quasar spectra; \citealt{2016ApJ...833..259B}). It is significantly more complicated to trace baryons in other regions of the Cosmic Web at redshift $z<1$, mainly because of the development of regions with very low gas column density. 

\cite{1998ApJ...503..518F} showed that approximately half of the total baryon budget at redshift $z<0.5$ is currently observable in galaxies, groups, clusters and the neutral Intergalactic Medium (IGM). The observable phases include the Interstellar Medium (ISM) of galaxies, the CGM in galaxies and galaxy groups, the hot Intracluster Medium (ICM) of galaxy clusters and the cool and diffuse phase of the IGM. Despite recent developments that allowed detailed theoretical and observational characterization of the CGM detectable in absorption and emission in proximity of galaxies (analytical models; \citealt{2018ApJ...856....5Q}; idealised Eulerian simulations; \citealt{2017MNRAS.466.3810F}; cosmological adaptive mesh refinement simulations; \citealt{2012MNRAS.420.1731F,2016ApJ...827..148C,2016MNRAS.458.1164L}; cosmological moving mesh simulations; \citealt{2013ApJ...772...97B,2015ApJ...804...72B,2017MNRAS.465.2966S,2018MNRAS.475.1160H}; cosmological smoothed particle hydrodynamics simulations; \citealt{2016MNRAS.459..310R,2016MNRAS.461L..32F}; Monte-Carlo radiative transfer methods; \citealt{2017ApJ...849..105L}), almost half of the cosmic baryon budget still defies direct detection. Over the last two decades, the analysis of cosmological hydrodynamical simulations has shown that a large fraction of these ``{\it missing} baryons''\footnote{The ``{\it missing} baryons'' problem we discuss in this paper is related to the distribution of baryons on cosmological scales. A separate ``{\it missing} baryons'' problem related to the baryonic mass content of individual halos has also been discussed in the literature \citep[e.g.][]{2007ARA&A..45..221B}. The latter is not the subject of our paper.} may reside in a shock-heated phase of the IGM with gas densities $\rho\lesssim 50\rho_{\rm crit}\Omega_{\rm b}$ and temperatures $ 10^5 \, {\rm K} < T < 10^7 \, {\rm K}$ \citep{1999ApJ...514....1C, 1999ApJ...511..521D, 2001ApJ...559L...5C, 2001ApJ...552..473D, 2005ApJ...620...21K, 2006ApJ...650..560C, 2007MNRAS.374..427D}; such a phase is usually denominated Warm-Hot Intergalactic Medium (WHIM). Based on these results from simulations, a variety of methods to directly detect the {\it missing} baryons has been proposed in the literature, e.g., via future radio surveys \citep{2012MNRAS.423.2325A, 2016arXiv160207526V, 2017PASJ...69...73H, 2018MNRAS.473...68C}, UV spectroscopy \citep{2009MNRAS.395.1875O, 2010MNRAS.408.1120B, 2011MNRAS.413..190T, 2018MNRAS.477..450N} and X-ray spectroscopy \citep{2001ApJ...554L...9P, 2003PASJ...55..879Y, 2005MNRAS.360.1110V, 2009ApJ...697..328B, 2011ApJ...734...91T, 2013arXiv1306.2324K}. 

Observational results show that UV OVI absorption lines from the WHIM can only trace a small fraction of the total baryon budget \citep{2008ApJ...679..194D, 2011ApJ...740...91P, 2012ApJ...759...23S, 2016ApJ...817..111D}. Statistically significant direct detection of WHIM in filaments using X-ray spectroscopy has been proven to be challenging \citep{2005Natur.433..495N, 2006ApJ...652..189K, 2007ApJ...656..129R, 2008Sci...319...55N, 2010ApJ...717...74Z}, but has recently proven to be fruitful \cite{2018Natur.558..406N}. 
Recent work has shown that it is possible to successfully identify baryonic filaments at redshift $z<0.5$ by stacking Lyman-$\alpha$ emission \citep{2018MNRAS.475.3854G} or the thermal Sunyaev-Zel'dovich signal \citep{2017arXiv170910378D,2017arXiv170905024T} from multiple filaments. HST/COS has also been successfully used to detect warm-hot gas associated with galaxy cluster pairs which has been interpreted as a promising route to detect the WHIM \citep{2016MNRAS.455.2662T}. 

Despite the progress, the bulk of the {\it missing} baryons is still extremely hard to detect and an analysis of the baryonic matter distribution in the Cosmic Web at low redshift has yet to be performed. The advent of the next generation of X-ray and radio telescopes may provide the tools to make significant breakthroughs in the quest for reconstructing the baryonic Cosmic Web \citep[see][]{2013arXiv1306.2324K, 2017PASJ...69...73H, 2018MNRAS.473...68C, 2016arXiv160207526V}. For this reason, providing solid predictions for the baryonic phases in the Cosmic Web at low redshift is very important to develop new observational strategies for future experiments. Furthermore, linking the physical state of baryons to their location in the Cosmic Web will allow us to make predictions relevant for available and upcoming infrared, optical and UV surveys. 

This paper is the first of a series focusing on the properties of baryons in the Cosmic Web. Here, we analyze the cosmological, hydrodynamical IllustrisTNG simulations and combine a detailed classification of cosmic structures with a characterization of different gas phases. We first identify cosmic structures, knots, filaments, sheets, and voids in the simulations, then classify the phases of baryonic gas residing within them. Such a characterization will be useful in future work to develop strategies to observe and measure properties of the baryonic Cosmic Web and its impact on galaxy evolution using the next generation of instruments \citep[see e.g.][]{2018MNRAS.474..547K}. Furthermore, IllustrisTNG is a suite of state-of-the-art cosmological, hydrodynamical simulations with significant improvements with respect to previous numerical work. Our analysis allows us to update and extend the results from previous authors that used simulations with lower resolution and earlier versions of popular numerical schemes and sub-resolution models. 

The paper is organised as follows: Section~\ref{sec:methods} provides an overview of the simulations and the analysis methods; Section~\ref{sec:results} shows our main results; Section~\ref{sec:discussion} discusses our results in the context of previous simulations and observations; Section~\ref{sec:summary} is dedicated to summary and conclusions.

\section{Methods}\label{sec:methods}

The purpose of this Section is to give an overview of the simulations we analyse (Subsection 2.1), of the Cosmic Web classification method we use (Subsection 2.2), of the conventions we adopt to define the phases of baryonic matter in the universe (Subsection 2.3), and of the criteria used to compute the column densities of several ions (Subsection 2.4).

\begin{figure*}
\begin{center}
 \includegraphics[width=0.99\textwidth]{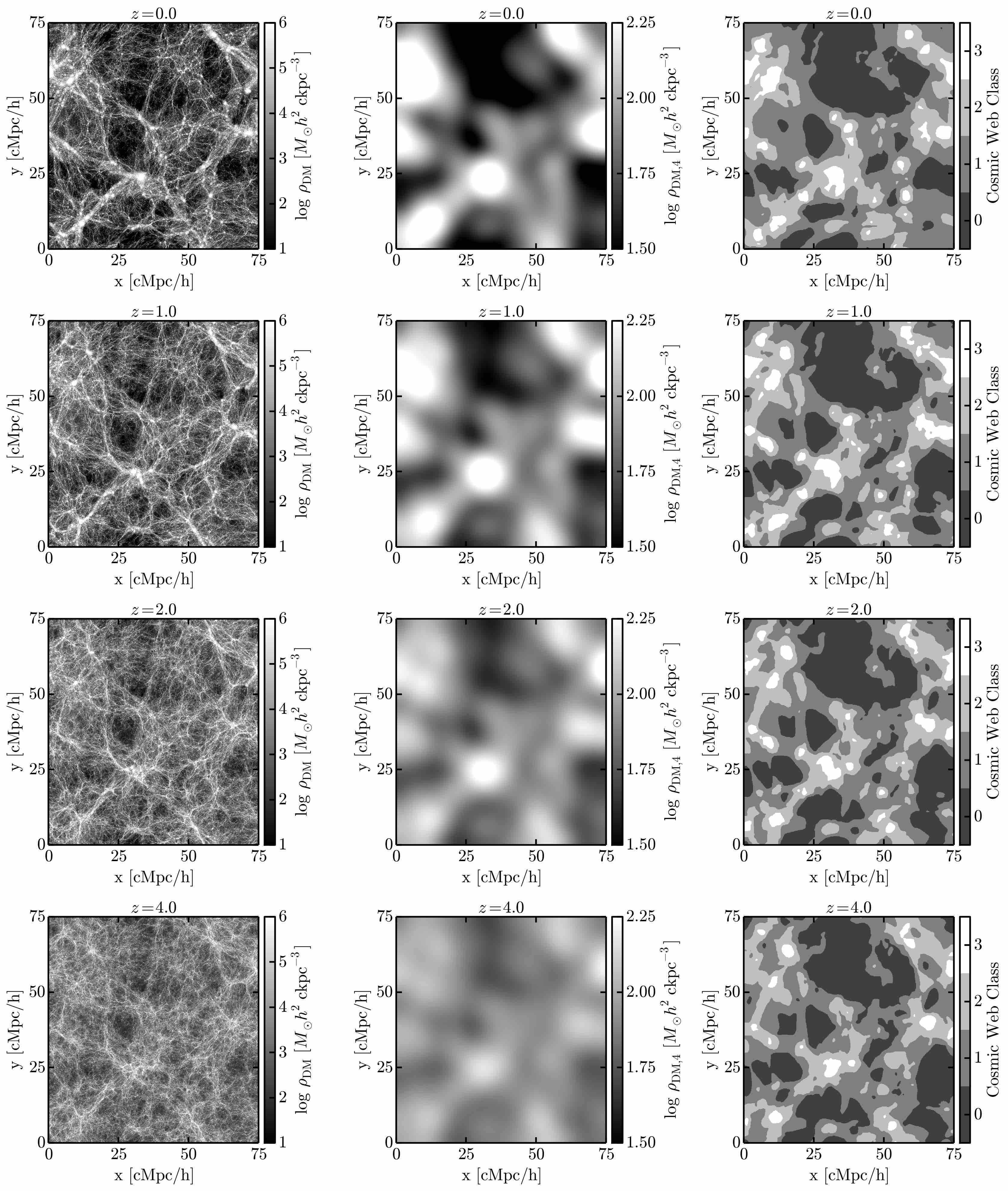}
\end{center}
\caption{ Left: dark matter mass density $\rho_{\rm DM}$ in a $75\times 75 \, {\rm cMpc^2}/h^2$ slice of thickness 8 ${\rm cMpc}/h$ passing through the box centre at various redshifts. Middle: dark matter density field smoothed with Gaussian kernel of radius $R_{\rm G} = 4 \, {\rm cMpc}/h$ in the same slice. Right: classification of the Cosmic Web in the same slices performed using our method with a Cartesian grid of size $512^3$ and Gaussian smoothing on a scale $R_{\rm G} = 4 {\rm cMpc}/h$; voids, sheets, filaments and knots are represented by progressively brighter shades of gray. Cosmic Web classes are defined by the eigenvalues of the deformation tensor: knots, filaments, sheets and voids have 3, 2, 1, 0 eigenvalues larger than $\lambda_{\rm th} =0.3$ (see Subsection 2.2), respectively. The Cosmic Web classification shown in the bottom panels is the average of the classification field along the line of sight for each pixel of the map computed only for cells in the thin slice we are considering. The algorithm yields satisfactory performance and $\sim$1\% precision in the determination of mass and volume fractions of each Cosmic Web component.}\label{fig:maps}
\end{figure*}

\subsection{IllustrisTNG}

Our results are based on the analysis of the TNG100 cosmological, hydrodynamical simulations \citep{2018MNRAS.475..676S, 2018MNRAS.480.5113M, 2018MNRAS.477.1206N, 2018MNRAS.473.4077P,2018MNRAS.475..648P, 2018MNRAS.475..624N} which are an updated version of the Illustris simulations \citep{2013MNRAS.436.3031V, 2014MNRAS.438.1985T, 2014MNRAS.444.1518V, 2014Natur.509..177V, 2014MNRAS.445..175G, 2015MNRAS.452..575S}. The IllustrisTNG model for galaxy formation~\citep[][]{2017MNRAS.465.3291W, 2018MNRAS.473.4077P} includes prescriptions for star formation, stellar evolution, chemical enrichment, primordial and metal-line cooling of the gas, stellar feedback with galactic outflows, black hole formation, growth and multimode feedback. The main differences between IllustrisTNG and Illustris are: (I) a new implementation of galactic winds, of which directionality, velocity, thermal content and energy scalings have been modified, (II) a new implementation of black-hole-driven kinetic feedback at low accretion rates, (III) and the inclusion of magnetohydrodynamics. Gas phase metal abundances in IllustrisTNG have been studied in \cite{2018MNRAS.474.2073V}, \cite{2018MNRAS.477..450N}, \cite{2017arXiv171105261T}, \cite{2018MNRAS.477L..16T}, \cite{2018MNRAS.477L..35G}. The IllustrisTNG dataset has been publicly released \citep{2018arXiv181205609N}.

In this paper, we use snapshots from the TNG100 simulations at redshift $0 \leq z\leq 8$. This simulation adopts a cubic box of length 75 cMpc/$h$ with periodic boundary conditions, assumes cosmological parameters from the \cite{2016A&A...594A..13P}, and follows the evolution of dark matter and baryons starting with $1820^3$ dark matter particles and $1820^3$ hydro cells. The dark matter particle mass is $m_{\rm dm} = 7.46 \times 10^6 \, {\rm M}_{\odot}$, whereas the initial baryonic mass particle is $ m_{\rm bar}=1.39 \times 10^6 \, {\rm M}_{\odot}$. The spatial resolution is $\sim 1$ kpc/$h$.

Since part of the results shown in this paper are related to the properties of the low density IGM, we correct for a small numerical error in the IllustrisTNG simulations which influences the value of the thermal energy in the lowest density regions. 
This feature is due to a code regression inadvertently introduced in an update of the velocity gradient computation, causing the Hubble flow across a fluid cell to be incorrectly taken into account in IllustrisTNG. As a result, some spurious numerical dissipation in expanding gas can arise, an effect that is only noticeable in cool gas of low density, where the cell sizes are comparatively large. We have verified that this only affects low density IGM gas in equilibrium between adiabatic cooling and photoheating from the background. This phase of the IGM should trace a power-law adiabat in the ($n_{\rm H},T$) diagram. However, in the snapshots of IllustrisTNG the adiabat has a slight upwards curvature at the lowest densities due to the numerical inaccuracy mentioned above.

In this paper, we use the updated values for the thermal energy of the fluid, that correct for this numerical feature, and that are provided by the IllustrisTNG public data release (\citealt[][for details on the correction]{2018arXiv181205609N}). The correction is only applied for gas cool gas ($T< 10^5 \, {\rm K}$) with physical hydrogen number density $n_{\rm H}< 10^{-6}(1+z)^{3} \, {\rm cm}^{-3}$, where the redshift-dependent factor has been introduced to take cosmological expansion into account. The details of the IGM temperature correction and its impact on the results of this paper are shown and quantified in Appendix~\ref{appendix:tcorr}. For our puroposes the effect of the correction is found to be negligible. 

\subsection{Cosmic Web Classification}\label{sec:class}

A number of different classification schemes with different levels of sophistication have been proposed in the literature to analyse the Cosmic Web based on the Hessian of the cosmic density field \citep{2007ApJ...655L...5A, 2007MNRAS.375..489H, 2009MNRAS.393..457S, 2009MNRAS.396.1815F, 2017ApJ...838...21Z,2018MNRAS.473...68C}, watershed segmentation of the cosmic density field \citep{2010ApJ...723..364A}, the velocity shear tensor \citep{2012MNRAS.425.2049H, 2016MNRAS.458.1517F, 2017ApJ...845...55P}, a combination of density and kinematic information \citep{2013MNRAS.429.1286C}, Bayesian reconstruction of the density field from a population of tracers \citep{2015A&A...576L..17L}, gradient-based methods that detects filaments through density ridges \citep{2016MNRAS.461.3896C}, network analysis \citep{2016MNRAS.459.2690H}, analysis of the dark matter flip-flop field \citep{2017MNRAS.468.4056S}, the identification of caustics \citep{2018JCAP...05..027F}, and Cosmic Web skeleton construction using Morse theory \citep{2013arXiv1302.6221S,2018MNRAS.481.4753C}. \cite{2018MNRAS.473.1195L} recently presented a thorough comparison of multiple methods. For this paper we choose to implement our own version of the method developed by \cite{2009MNRAS.396.1815F} which classifies the Cosmic Web using the deformation tensor and is based on results from the analysis of the growth of the large scale structure using the Zel'dovich approximation.

We briefly summarise the Cosmic Web classification method. We first interpolate the mass of each particle in IllustrisTNG to a Cartesian grid using a Cloud-In-Cell method to estimate the mass overdensity field in each cell, $\delta(\bf{x})$:
\begin{equation}
\delta(\bf{x})= \frac{\rho({\bf x})-\bar{\rho}}{\bar{\rho}},
\end{equation}
where $\rho({\bf x})$ is the matter density at position ${\bf x}$ and $\bar{\rho}$ is the mean matter density. We consider overdensity fields with Gaussian smoothing on scales $R_{\rm G} = 2$ and 4 cMpc/$h$. We classify each cell of the Cartesian grid as belonging to a knot, sheet, filament or void depending on the local eigenvalues of the deformation tensor $\Psi$, the Hessian of the gravitational potential $\phi$: $\Psi_{ij}({\bf x})=\partial_i \partial_j \phi({\bf x})$. Fast Fourier Transform (FFT) methods are used to compute the components of the deformation tensor at each point of the grid, using a method inspired by \cite{2007MNRAS.375..489H} and \cite{2009MNRAS.396.1815F}. The gravitational potential is related to the overdensity field via Poisson's equation 
\begin{equation}
\nabla^2\phi = 4\pi G \bar{\rho} \delta,
\end{equation}
By taking the FFT of the overdensity field $\delta({\bf x})$, we can compute the Fourier transform of the deformation tensor components as 
\begin{equation}
\Psi_{ij,{\bf k}} = k_i k_j \phi_{\bf k},
\end{equation}
which has been expressed in units where $4\pi G\bar{\rho} =1$, i.e. when the density unit ${\rho_{\rm unit}}$ is set in such a way that $\bar{\rho}=\rho_{\rm unit}/(4\pi G)$. By performing the inverse FFT of $\Psi_{ij,\bf{k}}$ for each $i,j$, we obtain the components of the deformation tensor at each point in space $\Psi_{ij}(\bf{x})$. For each position $\bf{x}$ on the Cartesian grid, we compute the eigenvalues of $\Psi(\bf{x})$, $\lambda_1(\bf{x}),\lambda_2(\bf{x}),\lambda_3(\bf{x})$ which are the solution of the following equation: 
\begin{equation}
\det({\bf \Psi}({\bf x})-{\it \lambda}({\bf x}){\bf I})= 0,
\end{equation}
where ${\rm \bf I}$ is the identity matrix. Classification is performed by counting the number of eigenvalues exceeding a threshold $\lambda_{\rm th}$: voids have all eigenvalues $<\lambda_{\rm th}$; sheets have one eigenvalue $\geq \lambda_{\rm th}$; filaments have two eigenvalues $\geq \lambda_{\rm th}$; knots have three eigenvalues $\geq \lambda_{\rm th}$.

The threshold $\lambda_{\rm th}$ is a free parameter and its value needs to be adjusted for different smoothing scales. \cite{2016ApJ...831..181L} set $\lambda_{\rm th} = 0.07$ to analyse the Cosmic Web in their N-body simulations, so that voids occupy a volume fraction $f_{\rm void}\sim 0.20$ at redshift $z=2.5$. \cite{2009MNRAS.396.1815F} suggest values in the range $0.2\leq \lambda_{\rm th} \leq 0.4$ that typically yield higher void volume fractions. We have experimented with different values in the range $0.07\leq \lambda_{\rm th} \leq 0.5$ (see Appendix~\ref{appendix:cwebtests}), and we ultimately settled to a fiducial value $\lambda_{\rm th}=0.3$, which yield values for volume and mass fraction of cosmic structure similar to those found by other authors (see Section~\ref{sec:webfracs}).

Figure~\ref{fig:maps} shows a visual comparison of density maps from the simulations and the result of the Cosmic Web classification algorithm at redshifts $z=0,1,2,4$. Figure~\ref{fig:maps} qualitatively highlights the success of the classification algorithm at reconstructing the overall structure of the Cosmic Web at multiple redshifts. 

We also perform extensive quantitative tests of the classification scheme. We vary the resolution of the Cartesian grid between $256^3$, $512^3$, $1024^3$ to assess the convergence of the algorithm as resolution is increased. We find that for grid sizes $\geq 512^3$ the mass and volume fractions of the different cosmic structures measured by the classification scheme do not vary more than $\sim 1\%$ as the resolution is further increased. We also vary the Gaussian smoothing scale and we do not find significant variations in the mass and volume fractions we measure. The results of our tests are in excellent agreement with the tests of the same method performed by \cite{2009MNRAS.396.1815F}. See Appendix~\ref{appendix:cwebtests} for details on our tests. Our fiducial parameters for the classification method are smoothing scale $R_{\rm G} = 4$ cMpc/$h$ and a Cartesian grid size $512^3$. 

% NEW TABLE WITH LAMBDA_TH = 0.3
\begin{table}
\centering
\caption{Mass and volume fractions of all gaseous phases of baryonic matter, $f_{\rm gas,M}$ and $f_{\rm gas,V}$, respectively. The mass fractions are normalised with respect to the total gas mass. Volume fractions are computed with respect to the total volume. Results are reported for  redshifts $z=0, 1, 2, 4$ and are achieved by integrating over all regions of the Cosmic Web, so the values do not depend on the classification algorithm. }\label{tab:phases}
{\bfseries Mass and Volume Fractions of Gas Phases}
\makebox[\linewidth]{
\begin{tabular}{llcc}
\hline
\hline
\multicolumn{4}{l}{Redshift $z=0.0$} \\ 
\hline
 & Phase & $f_{\rm gas,M}$ & $f_{\rm gas,V}$ \\
\hline
 & Star-forming Gas & $3.3\times 10^{-3}$ & $2.5\times 10^{-9}$ \\
 & Halo Gas & $4.3\times 10^{-2}$ & $9.9\times 10^{-6}$ \\
 & Diffuse IGM & $3.9\times 10^{-1}$ & $8.9\times 10^{-1}$ \\
 & WHIM & $4.6\times 10^{-1}$ & $1.1\times 10^{-1}$ \\
 & WCGM  & $3.1\times 10^{-2}$ & $3.6\times 10^{-5}$ \\
 & HM  & $7.3\times 10^{-2}$ & $1.0\times 10^{-3}$  \\
\hline
\hline
\multicolumn{4}{l}{Redshift $z=1.0$} \\ 
\hline
 & Phase & $f_{\rm gas,M}$ & $f_{\rm gas,V}$ \\
\hline
 & Star-forming Gas & $1.0\times 10^{-2}$ & $3.9\times 10^{-8}$ \\
 & Halo Gas & $6.3\times 10^{-2}$ & $6.9\times 10^{-5}$  \\
 & Diffuse IGM & $5.8\times 10^{-1}$ & $9.6\times 10^{-1}$  \\
 & WHIM & $2.6\times 10^{-1}$ & $3.9\times 10^{-2}$ \\
 & WCGM  & $7.2\times 10^{-2}$ & $2.2\times 10^{-4}$ \\
 & HM  & $1.6\times 10^{-2}$ & $2.1\times 10^{-4}$  \\
 \hline
\hline
\multicolumn{4}{l}{Redshift $z=2.0$} \\ 
\hline
 & Phase & $f_{\rm gas,M}$ & $f_{\rm gas,V}$ \\
\hline
 & Star-forming Gas & $1.3\times 10^{-2}$ & $1.6\times 10^{-7}$ \\
 & Halo Gas & $9.6\times 10^{-2}$ & $3.9\times 10^{-4}$  \\
 & Diffuse IGM & $7.1\times 10^{-1}$ & $9.8\times 10^{-1}$  \\
 & WHIM & $1.1\times 10^{-1}$ & $1.8\times 10^{-2}$ \\
 & WCGM  & $6.5\times 10^{-2}$ & $4.1\times 10^{-4}$ \\
 & HM  & $2.3\times 10^{-3}$ & $7.1\times 10^{-5}$  \\
\hline
\hline
\multicolumn{4}{l}{Redshift $z=4.0$} \\ 
\hline
 & Phase & $f_{\rm gas,M}$ & $f_{\rm gas,V}$ \\
\hline
 & Star-forming Gas & $8.0\times 10^{-3}$ & $5.5\times 10^{-7}$ \\
 & Halo Gas & $1.5\times 10^{-1}$ & $2.4\times 10^{-3}$  \\
 & Diffuse IGM & $7.9\times 10^{-1}$ & $9.9\times 10^{-1}$  \\
 & WHIM & $3.2\times 10^{-2}$ & $1.2\times 10^{-2}$ \\
 & WCGM  & $1.7\times 10^{-2}$ & $3.0\times 10^{-4}$ \\
 & HM  & $3.8\times 10^{-5}$ & $3.5\times 10^{-7}$  \\
\hline
\hline
\end{tabular}
}
\end{table}

%%%%%%%%%%%%%%%%%%%%%%

\subsection{Gas phases}\label{sec:phases}

After the Cosmic Web classification has been performed, we group the baryonic particles in IllustrisTNG into several phases which have different observational properties. We compute mass and volume fractions of each of the following gas phases:
\begin{itemize}
\item Star-forming Gas: number density $n_{\rm H} > 0.13 \, {\rm cm^{-3}}$, temperature $T < 10^7 \, {\rm K}$ and star formation rate ${\rm SFR > 0}$. Gas with $n_{\rm H} > 0.13 \, {\rm cm^{-3}}$ is identified as star-forming in the IllustrisTNG model, but we place an additional temperature cut in post-processing. The star formation rate cut selects only gas that is effectively forming stars.
\item Halo Gas: number density $10^{-4}(1+z) \, {\rm cm^{-3}} < n_{\rm H} < 0.13 \, {\rm cm^{-3}}$ and temperature $T < 10^5 \, {\rm K}$. This phase contains `cool' gas associated with the ISM of galaxies and with the CGM observed in the halos of galaxies. Since both types of gas are found inside or in the proximity of galaxies, and they have similar temperature and densities, we group them together.
\item Diffuse IGM: number density $n_{\rm H} < 10^{-4}(1+z) \, {\rm cm^{-3}}$ and temperature $T < 10^5 \, {\rm K}$. At redshift $z=0$ the number density cutoff corresponds to regions of gas density $\rho<25\rho_{\rm crit}\Omega_{\rm b}$, where $\rho_{\rm crit}$ is the critical density. The density cutoff selects gas at small to mild overdensities, effectively capturing the typical range associated with intergalactic gas. The redshift evolution takes into account the fact that the gas in this phase is located in regions whose over/underdensity evolves in the linear regime. This combination of density and temperature cutoffs  selects the phase of the IGM in which hydrogen can be either neutral or ionized, but helium and heavier elements are not completely ionized. 
\item WHIM: number density $n_{\rm H} < 10^{-4}(1+z) \, {\rm cm^{-3}}$ and temperature $10^5 \, {\rm K} < T < 10^7 \, {\rm K}$. This phase has the same characteristic density of the Diffuse IGM, but the temperature cutoff is such that gas can contain a significant abundance of ionized helium and heavier elements. 
\item WCGM: number density $10^{-4}(1+z) \, {\rm cm^{-3}} < n_{\rm H} < 0.13  \, {\rm cm^{-3}}$ and temperature $10^5 \, {\rm K} < T < 10^7 \, {\rm K}$. This gas exists at temperatures comparable to the WHIM, but is located in regions with higher overdensity $\rho>25\rho_{\rm crit}\Omega_{\rm b}$. Dense gas in this temperature range is more efficiently created by shock heating and feedback processes near galaxies. For this reason, we expect the cutoffs to select gas in the warm CGM of galaxies and galaxy groups, hence the `WCGM' label. 
\item Hot Medium (HM): any number density ${\rm n_H}$ and temperature $T > 10^7 {\rm K}$. The cutoffs selects gas with temperature larger than the virial temperature of massive galaxy clusters ($M_{\rm vir}\gtrsim 10^{14}$ M$_{\odot}$). For this reason, the cutoffs are selecting gas that has been shock heated to these high temperatures in (and near) the most massive dark matter halos in the universe.
\end{itemize}
These baryonic phases will be referenced multiple times in the text below. In particular, we will provide mass and volume fractions of each phase in different regions of the Cosmic Web, which will facilitate comparison to results from other cosmological simulations and to measurements of the abundances from observations \citep[e.g.][]{1998ApJ...503..518F}. \cite{2016MNRAS.457.3024H} performed a similar study of baryonic phases using the original Illustris simulation, but their definitions do not distinguish between non-star-forming and star-forming gas, and between WHIM and WCGM. We find that this distinction is important because these two phases have different evolution with redshift (Section~\ref{sec:results}). We discuss the differences between our definitions and the ones used by \cite{2016MNRAS.457.3024H} in Section~\ref{sec:discussion}. 

\subsection{Ion Densities}\label{sec:ions}

With the goal of comparing to observations, we analyse IllustrisTNG to measure the contribution to the column densities of OVII and NeIX from different regions of the Cosmic Web. To achieve this goal, we use the cloud-in-cell mass assignment scheme to compute the number density of each ion in the cells of a $512^3$ mesh identical to that used for the Cosmic Web classification, which allows us to perform a joint analysis of the ionization state and of the location in the Cosmic Web. We perform ionization modeling generating ionization tables that depend on density, temperature and metallicity with the {\sc cloudy} code \citep{2013RMxAA..49..137F} assuming the UV background of \cite{2009ApJ...703.1416F}. For each gas particle, we use the local density, temperature and metallicity to retrieve the appropriate ionization state from the {\sc Cloudy} tables, which is then used to compute the ion density. We use 8 equally spaced logarithmic bins for the density in the range ${-8\leq \log(n_{\rm H}/{\rm cm^{-3}}) \leq 4}$, 14 equally spaced logarithmic bins for the temperature in the range ${2\leq \log(T/{\rm K}) \leq 9}$, and 20 equally spaced metallicity bins in the range ${-4\leq \log(Z/{\rm Z_{\odot}}) \leq 1}$. Our procedure is similar to that used by \cite{2018MNRAS.477..450N}, but they use finer spacing in density and temperature, and coarser spacing in metallicity. Since we focus on absorption lines from the low density IGM, we do not include self-shielding of high density gas. We note that gas phase metal abundances, especially in low temperature or warm regions, can be reduced due to dust condensation~\citep[][]{McKinnon2016,McKinnon2017}, which we do not model here. 

\begin{figure*}
\begin{center}
 \includegraphics[width=0.99\textwidth]{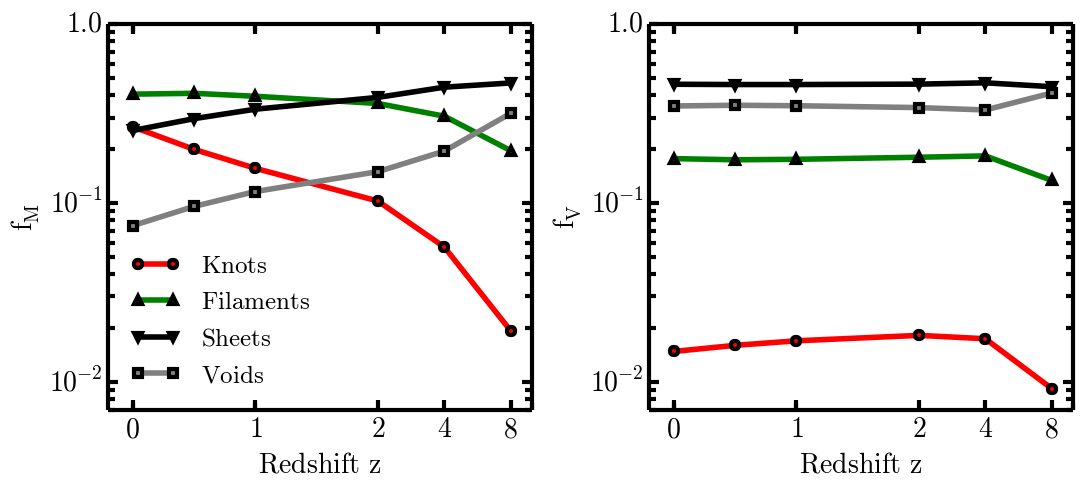}
\end{center}
\caption{ Left: evolution of the total (dark matter + baryonic) mass fraction of different cosmic structures with respect to the total dark matter and baryonic mass in the simulated volume from redshift $z=8$ to redshift $z=0$. Right: evolution of the volume fractions of different cosmic structures with respect to the total simulated volume from $z=8$ to redshift $z=0$. Most of the mass in the universe is in knots, filaments and sheets. Most of the volume is occupied by sheets, filaments and voids. The mass fraction of knots and voids evolve dramatically from high to low redshift, whereas filaments and sheets evolve weakly.}\label{fig:web_fractions_evo}
\end{figure*}

% NEW TABLE FOR LAMBDA_TH = 0.3
\begin{table*}
\centering
\caption{Mass and volume fractions of different structures in the Cosmic Web, $f_{\rm M}$ and $f_{\rm V}$, respectively. Both baryonic and dark matter are included. The reported values are measured adopting a Cartesian grid of size $512^3$ for the classification, and two Gaussian smoothing scales $R_{\rm G} = 2$ and 4 cMpc/$h$. Only significant digits that are robust against variation of grid size between $256^3$, $512^3$ and $1024^3$ are reported.}\label{tab:fractions}
{\bfseries Dark Matter + Baryons Mass and Volume Fractions of Cosmic Structures}
\makebox[\linewidth]{
\begin{tabular}{llcccc}
\hline
\hline
\multicolumn{6}{l}{Redshift $z=0$} \\ 
\hline
 & Type & $f_{\rm M} (R_{\rm G}=2 \, {\rm cMpc}/h)$ & $f_{\rm M} (R_{\rm G}=4 \, {\rm cMpc}/h)$ & $f_{\rm V} (R_{\rm G}=2 \, {\rm cMpc}/h)$ & $f_{\rm V} (R_{\rm G}=4 \, {\rm cMpc}/h)$ \\
\hline
 & Knots & 0.33 & 0.27 & 0.01 & 0.01 \\
 & Filaments & 0.40 & 0.40 & 0.16 & 0.18 \\
 & Sheets & 0.21 & 0.25 & 0.46 & 0.46 \\
 & Voids & 0.06 & 0.08 & 0.37 & 0.35 \\
\hline
\hline
\multicolumn{6}{l}{Redshift $z=1$} \\ 
\hline
 & Type & $f_{\rm M} (R_{\rm G}=2 \, {\rm cMpc}/h)$ & $f_{\rm M} (R_{\rm G}=4 \, {\rm cMpc}/h)$ & $f_{\rm V} (R_{\rm G}=2 \, {\rm cMpc}/h)$ & $f_{\rm V} (R_{\rm G}=4 \, {\rm cMpc}/h)$ \\
\hline
 & Knots & 0.21 & 0.16 & 0.01 & 0.02 \\
 & Filaments & 0.40 & 0.39 & 0.16 & 0.17 \\
 & Sheets & 0.30 & 0.33 & 0.46 & 0.46 \\
 & Voids & 0.09 & 0.12 & 0.37 & 0.35 \\
\hline
\hline
 \multicolumn{6}{l}{Redshift $z=2$} \\ 
\hline
 & Type & $f_{\rm M} (R_{\rm G}=2 \, {\rm cMpc}/h)$ & $f_{\rm M} (R_{\rm G}=4 \, {\rm cMpc}/h)$ & $f_{\rm V} (R_{\rm G}=2 \, {\rm cMpc}/h)$ & $f_{\rm V} (R_{\rm G}=4 \, {\rm cMpc}/h)$ \\
\hline
 & Knots & 0.14 & 0.10 & 0.02 & 0.02 \\
 & Filaments & 0.38 & 0.36 & 0.16 & 0.18 \\
 & Sheets & 0.35 & 0.39 & 0.46 & 0.46 \\
 & Voids & 0.13 & 0.15 & 0.36 & 0.34 \\
\hline
\hline
 \multicolumn{6}{l}{Redshift $z=4$} \\ 
\hline
 & Type & $f_{\rm M} (R_{\rm G}=2 \, {\rm cMpc}/h)$ & $f_{\rm M} (R_{\rm G}=4 \, {\rm cMpc}/h)$ & $f_{\rm V} (R_{\rm G}=2 \, {\rm cMpc}/h)$ & $f_{\rm V} (R_{\rm G}=4 \, {\rm cMpc}/h)$ \\
\hline
 & Knots & 0.08 & 0.06 & 0.02 & 0.02 \\
 & Filaments & 0.33 & 0.31 & 0.17 & 0.18 \\
 & Sheets & 0.42 & 0.44 & 0.47 & 0.47 \\
 & Voids & 0.17 & 0.19 & 0.34 & 0.33 \\
\hline
\hline
\end{tabular}
}
\end{table*}

\section{Results}\label{sec:results}

We focus our analysis on the redshift range $0\leq z \leq 8$ which covers both the search for the {\it missing} baryons ($0 \leq z \leq 1$) and the redshift evolution of the baryons. Before discussing the details of our joint analysis of cosmic structures and baryonic phases, we report the total mass and volume fraction of all the considered baryonic phases at redshift $z=0, 1, 2, 4$ (Table~\ref{tab:phases}). The mass fractions are normalised with respect to the total gas mass throughout the whole Section. The same convention was used by \cite{2017arXiv171105261T} in their independent analysis of gas phases and metallicities in IllustrisTNG. 

At redshift $z=0$ Star-forming Gas and Halo Gas always constitute a small fraction of the total baryon budget ($f_{\rm gas,M}<5\times 10^{-2}$) and occupy a small volume, since this phase is only found in galaxies.  The total contribution from the WCGM is only a $\sim 3$\% of the total baryon budget and its volume fraction is $\sim 4 \times 10^{-5}$. HM constitutes $\sim 7.5\%$ of the total baryon budget at redshift $z=0$, but it only occupies $\sim 0.1\%$ of the total volume. Indubitably, the two dominant components of the baryonic Cosmic Web at redshift $z=0$ are the Diffuse IGM and the WHIM, which constitute $\sim 38\%$ and $\sim 47\%$ of the baryonic mass at redshift $z=0$, respectively, and occupy most of the volume. 

The mass fraction of HM decreases by a factor $\sim 4$ from redshift $z=0$ to redshift $z=1$ and by a factor $\sim 400$ from redshift $z=1$ to $z=4$, as a result of the lower abundance of massive halos able to shock heat gas to high temperatures at higher redshift. WCGM increases out to redshift $z=1$, but declines dramatically at redshift $z>1$. The WHIM mass/volume fraction decreases by a factor $\sim 2$ from redshift $z=0$ to redshift $z=1$, whereas the Diffuse IGM mass fraction increases. The abundance of the WHIM decreases by a factor $\sim 10$ from $z=1$ to $z=4$, whereas the abundance of Diffuse IGM keeps increasing with redshift. These results suggest that the WHIM should be a significant fraction of the baryon budget even at redshift $z=1$, but not at higher redshift where most of the baryonic mass (and volume) in the Cosmic Web is represented by Diffuse IGM.

Table~\ref{tab:phases} shows how much baryonic mass is present in each phase at each redshift, but it does not tell {\it where} each phase lies in the Cosmic Web. Details of the evolution of baryonic phases in each class of cosmic structure will be discussed in the next Subsections.

\subsection{Mass and Volume Fractions of Cosmic Structures}\label{sec:webfracs}

Before re-analysing baryonic phases, we first focus on the total (dark and baryonic) mass budget in the Cosmic Web. Figure~\ref{fig:web_fractions_evo} shows the evolution of the mass and volume fractions of knots, filaments, sheets and voids in the redshift range $0\leq z \leq 8$. Numerical values for these fractions are summarized in Table~\ref{tab:fractions} for redshift $z=0, 1, 2, 4$. Figure~\ref{fig:web_fractions_evo} shows that filaments and sheets dominate the cosmic budget of all matter at all redshifts, both in terms of mass and volume fractions. Knots accumulate more mass at redshift $z<1$, where increasingly massive dark matter halos form. The increase of mass in knots is matched by a decrease in mass in sheets over the whole redshift range $0\leq z \leq 8$. The mass fraction of filaments doubles from redshift $z=8$ to redshift $z=0$. Finally, the mass fraction of voids steadily decreases from high to low redshift. Figure~\ref{fig:web_fractions_evo} also shows a weak redshift evolution of the volume fractions of each class of cosmic structures. 

\begin{figure*}
\begin{center}
 \includegraphics[width=0.49\textwidth]{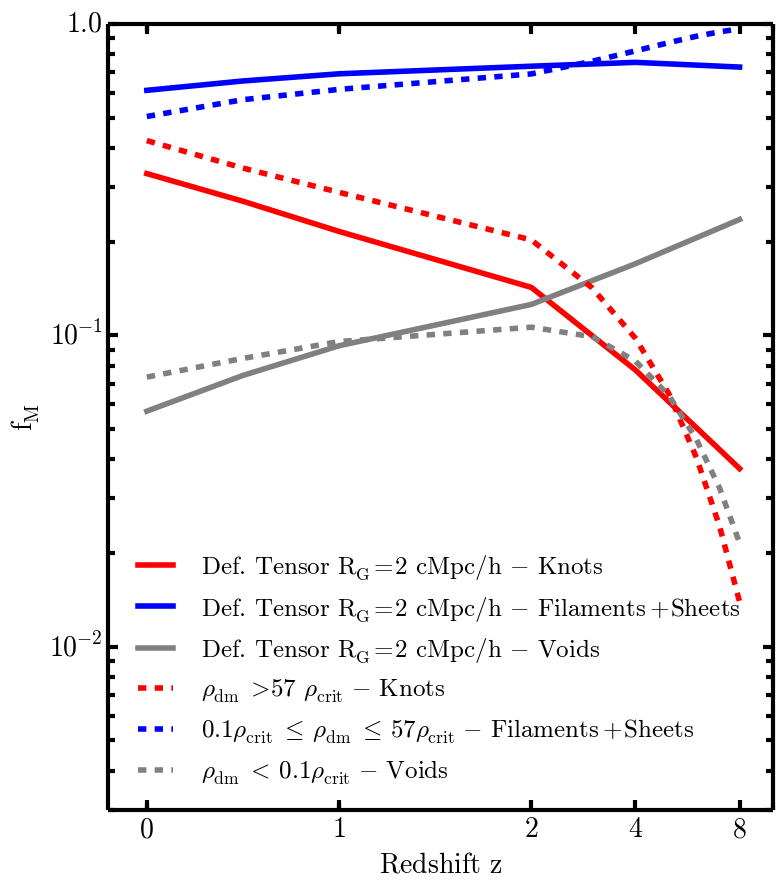}
 \includegraphics[width=0.49\textwidth]{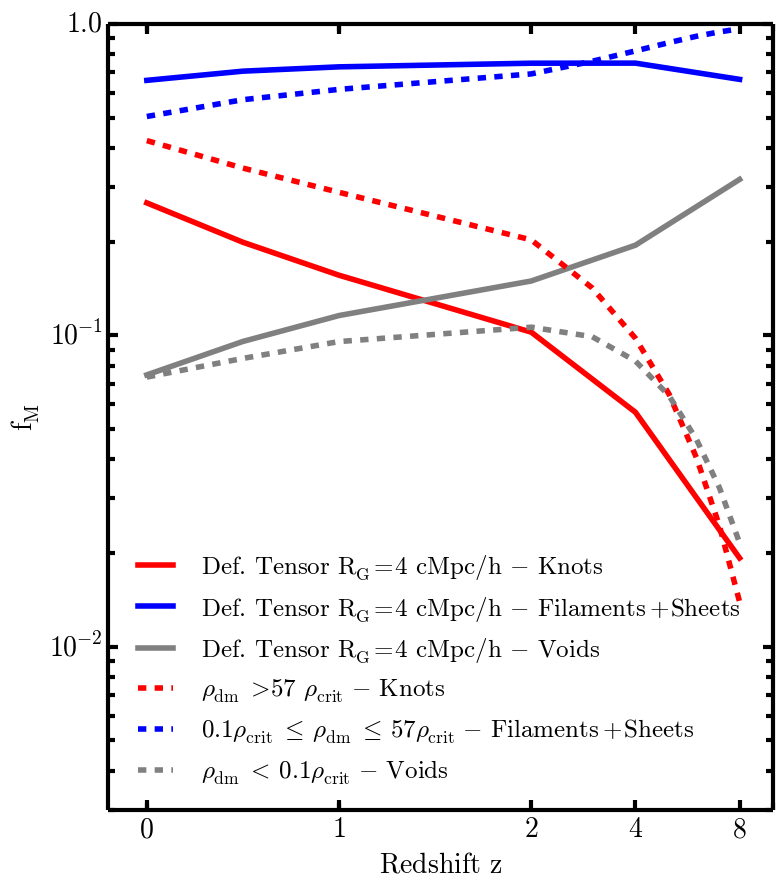}
\end{center}
\caption{Evolution of the total (dark matter + baryonic) mass fraction of different cosmic structures with respect to the total dark matter and baryonic mass in the simulated volume from redshift $z=8$ to redshift $z=0$. Two methods for the Cosmic Web classification are compared: (I) our fiducial method based on the deformation tensor (\citealt{2009MNRAS.396.1815F}; solid lines), and (II) a method which uses local dark matter density cuts (\citealt{2016MNRAS.457.3024H}; dashed lines). Left: the deformation tensor method uses a density field smoothed using a Gaussian kernel of radius $R_{\rm G}=2$ cMpc/$h$. Right: the deformation tensor method uses a density field smoothed using a Gaussian kernel of radius $R_{\rm G}=4$ cMpc/$h$. Differences between the two classification methods are maximal at redshift $z>4$ because the density at these redshift is $\rho_{\rm dm}\sim \Omega_{\rm m}\rho_{\rm crit}>0.1\rho_{\rm crit}$ and density fluctuations are small, so the local density cut method will assign most regions in the volume to filaments+sheets or knots. At redshift $z<4$ the two methods produce similar trends in the mass fraction, but different exact values. This is a consequence of density field smoothing. The method based on density cuts only smoothes the density field over an SPH kernel, whereas the deformation tensor method uses smoothing on much larger scales. As a consequence, the difference between the two methods is minimised by decreasing the smoothing length of the density field used in the deformation tensor method. }\label{fig:web_fractions_evo_2methods}
\end{figure*}

To assess the robustness of our Cosmic Web classification method, we compare it to an alternative method similar to the one used by \cite{2016MNRAS.457.3024H} to analyse the original Illustris simulations, that will also be used in Artale et al. (in prep.). This method uses the local dark matter density estimated as the standard cubic-spline SPH kernel over all the dark matter particles within a certain radius. In this alternative method, knots are defined as those regions where the local dark matter density is higher than $57\rho_{\rm crit}$. Filaments and sheets are defined as the regions with local dark matter density between $0.1 - 57 \rho_{\rm crit}$. Finally, voids are defined as regions with local dark matter density lower than $0.1\rho_{\rm crit}$. The biggest difference of this method with respect to our fiducial one based on the deformation tensor is that the density field used to detect cosmic structure is not smoothed, i.e. it is susceptible to small scale fluctuations. Figure~\ref{fig:web_fractions_evo_2methods} highlights differences between the two methods. At redshift $z>4$ the density cut method yields solutions that are very different from the deformation tensor method. At redshift $z>4$, the typical density is $\rho_{\rm dm}\sim\Omega_{\rm m}\rho_{\rm crit}> 0.1 \rho_{\rm crit}$ and density fluctuations are small. Therefore, in the method using local density cuts, most material will be assigned to filaments, sheets and knots. This effect is not present for the deformation tensor method which uses derivatives of the density field (contained in the deformation tensor; equations 1, 2 and 3), rather than the face value of the density field. At redshift $z<4$ the discrepancy between the two methods is alleviated and the mass fractions of baryons and dark matter of different cosmic structures with respect to the total mass exhibit similar trends. However, the exact values of these mass fractions at redshift $z<4$ depend on the smoothing of the density field. In particular, choosing a smaller smoothing length for the deformation tensor method produces results more similar to the local density cut method. The structures that are more susceptible to variation of the smoothing length are voids. The differences can be explained by considering that the method using local density cuts is based on a density field smoothed with the local SPH kernel whose width is much smaller than the smoothing scales we adopt in the deformation tensor method. As the smoothing scale is decreased, the deformation tensor method will converge to the same solution as the local density cut method.

As shown by \cite{2018MNRAS.473.1195L}, the exact values of the mass and volume fractions assigned to each cosmic structure depend on the classification method adopted for the task. We find that the differences in the mass fractions measured by the two methods that we consider in this subsection are not larger than the spread of the same quantities measured by the twelve algorithms used by \cite{2018MNRAS.473.1195L}. \cite{2009MNRAS.396.1815F} used the deformation tensor method on dark matter only simulations. For the same classification threshold $\lambda_{\rm th}$ and smoothing length $R_{\rm G}$ (see Subsection~\ref{sec:class}), comparison to \cite{2009MNRAS.396.1815F} indicates that the mass fraction of knots at redshift $z=0$ in IllustrisTNG is higher by a factor $\sim 2$ with respect to that in dark matter only simulations. On the other hand, the mass fraction of filaments in IllustrisTNG is similar to that of the dark matter simulations of \cite{2009MNRAS.396.1815F}. The mass fraction of sheets in IllustrisTNG is $\sim 2/3$ of that found by \cite{2009MNRAS.396.1815F}. Finally, the mass fraction of voids in IllustrisTNG is a factor $\sim 2$ smaller than that of \cite{2009MNRAS.396.1815F}. We do not expect a perfect quantitative match with the results of \cite{2009MNRAS.396.1815F}, because they do not include baryons and assume different cosmological parameters from WMAP3, whereas we include baryons and assume Planck 2015 parameters. Nonetheless, our results are in qualitative agreement with \cite{2009MNRAS.396.1815F}. 

\cite{2017ApJ...838...21Z} show the evolution of mass and volume fractions of the four cosmic structures identified with the same method with $\lambda_{\rm th} = 0.2$ and $0.4$, which bracket our fiducial value $\lambda_{\rm th}=0.3$, but they adopt smaller smoothing lengths $R_{\rm G}\geq 390$ kpc/h. Furthermore, the simulations of \cite{2017ApJ...838...21Z} include baryons, but they use a different set of cosmological parameters (WMAP5). These differences influence the exact values of the mass and volume fractions of each structure. The weak redshift evolution of the volume fractions of sheets, filaments and voids shown by \cite{2017ApJ...838...21Z} is in good agreement with our findings. The strong evolution of the mass fraction of voids and knots found in IllustrisTNG is also in agreement with the evolution reported in \cite{2017ApJ...838...21Z}. Given these considerations, we conclude that our fiducial method based on the deformation tensor computed from the density field smoothed with a Gaussian kernel of radius $R_{\rm G}=4$ Mpc/$h$ is adequate to characterise the large scale distribution of matter in the Cosmic Web and that our implementation performs similarly to implementations discussed in previous literature.

Finally, the limited size of TNG100 may in principle influence our results on the mass fraction of cosmic structures. In Appendix~\ref{appendix:boxsize} we show a comparison of the mass fractions of each cosmic structure in TNG100 and TNG300. The latter uses the IllustrisTNG model but follows cosmic structure formation in a volume $\sim 20$ times larger than TNG100. Our tests show that the mass fraction of associated to each cosmic structure is well converged when the box size is increased (Appendix~\ref{appendix:boxsize}).

\subsection{Baryonic Phase Diagram in Different Cosmic Structures}

\begin{figure*}
\begin{center}
 \includegraphics[width=0.99\textwidth]{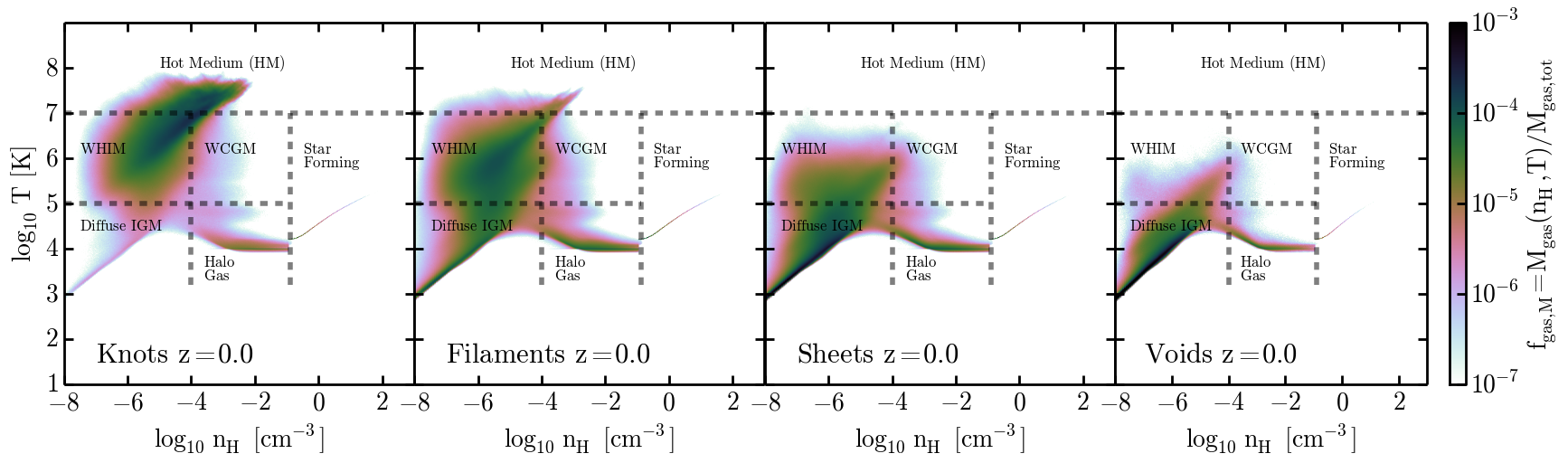}
 \includegraphics[width=0.99\textwidth]{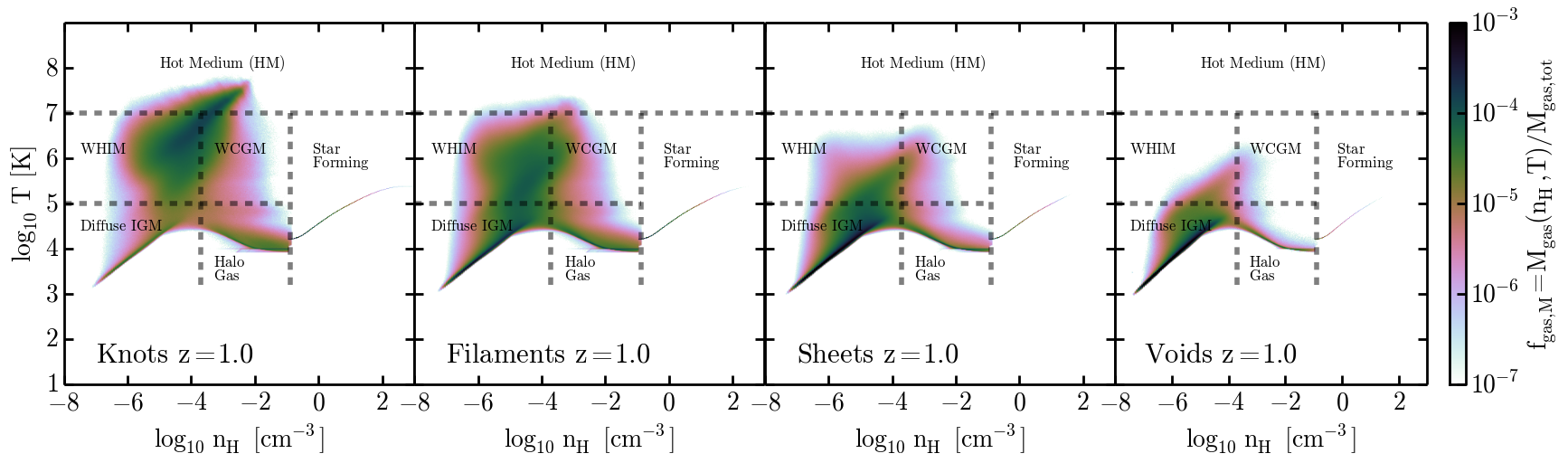}
 \includegraphics[width=0.99\textwidth]{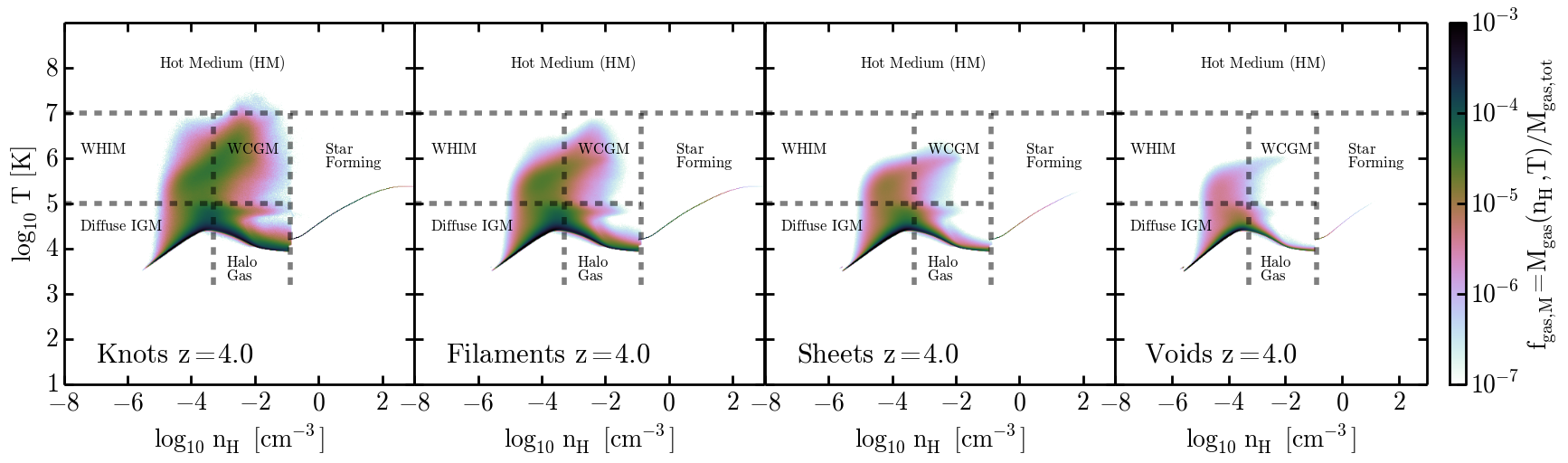}
\end{center}
\caption{ Phase diagram of baryons at redshift $z=0, 1$, and 4 in different regions of the Cosmic Web. Our classification method with Cartesian grid of size $512^3$ and Gaussian smoothing on a scale $R_{\rm G}=4$~cMpc/$h$ has been used to produce this plot. The phase diagram of baryons depends significantly on the location in the Cosmic Web. In particular, the hot and warm phase are only abundant in knots and filaments at redshift $z<1$.}\label{fig:mhist}
\end{figure*}

Figure~\ref{fig:mhist} shows 2-d histograms of the mass fraction of baryons with respect to the total baryonic mass in the density-temperature plane. We make a separate plot for the phase diagram baryons residing in knots, filaments, sheets and voids, respectively. We find significant differences between the phase diagrams of baryons residing in different cosmic structures. At redshift $z=0$ knots and filaments contain almost all warm-hot ($10^5 \, {\rm K} < T < 10^7 \, {\rm K}$) and hot gas ($T>10^7\, {\rm K}$) in the universe as a result of shock heating. The hot phase is almost completely absent in sheets and voids. Filaments, sheets and voids contain $\sim 90\%$ of all the diffuse phase of the IGM ($n_{\rm H} < 10^{-4} \, {\rm cm^{-3}}$ and $T < 10^5 \, {\rm K}$). At redshift $z=1$ we see a similar trend moving from knots to voids, but the warm-hot and hot phases are less prominent in all regions of the Cosmic Web. At redshift $z=4$ the hot phase is only present in knots and constitutes only $\sim 3.6\times10^{-3}\%$ of the gas in the universe. More than $60\%$ of the gas in the universe at redshift $z=4$ is cool (${T<10^5}$ K). 

\subsection{Probability Density Function of Baryonic Phases}\label{sec:pdf}

\begin{figure*}
\begin{center}
 \includegraphics[width=0.99\textwidth]{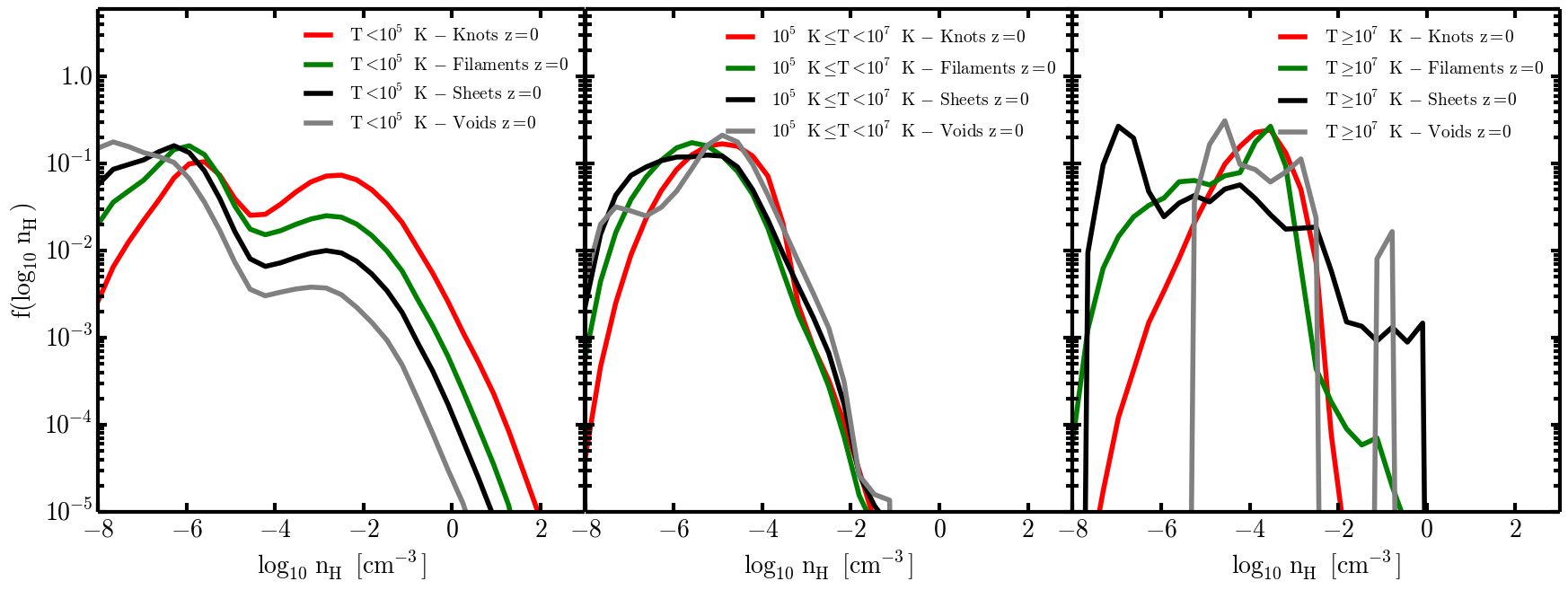}
\end{center}
\caption{ Density PDF as a function of the location in the Cosmic Web at redshifts $z=0$. Left panel: cool gas at temperature $T<10^5 \, {\rm K}$; since real star-forming gas is cold, the Star-forming Gas in the simulation has been added to this histogram, even if it formally lies on a polytropic equation of state. Middle panel: warm-hot gas at temperature $10^5 \, {\rm K} \leq T<10^7 \, {\rm K}$. Right panel: hot gas at temperature $T\geq 10^7 \, {\rm K}$. The density PDF demonstrates the existence of multiple baryonic phases and is significantly environment-dependent for the cool and hot gas.}\label{fig:pdf}
\end{figure*}

We investigate on the physical state of baryons in the Cosmic Web by computing probability density functions (PDFs) of the hydrogen number density in knots, filaments, sheets and voids and in different temperature ranges. Figure~\ref{fig:pdf} shows the PDFs at redshift $z=0$ for gas in the temperature ranges $T<10^5 \, {\rm K}$ (cool), $10^5 \, {\rm K} \leq T<10^7 \, {\rm K}$ (warm-hot), $T\geq 10^7 \, {\rm K}$ (hot), respectively. 

The left panel in Figure~\ref{fig:pdf} shows the density PDF of cool gas ($T<10^5 \, {\rm K}$) at redshifts $z=0$. In the simulation, Star-forming Gas lies on a polytropic equation of state, but real star-forming gas is cold. For this reason, we add the Star-forming gas in the simulation to the histogram of the cool gas. We find a bi-modal distribution as a function of density. The low density bump of the PDF is associated with the Diffuse IGM, whereas the high density peak is associated with the Halo Gas and Star-forming Gas. It is clear from this figure that filaments and knots tend to host much larger fractions of cool, dense gas than sheets and voids. 

The middle panel of Figure~\ref{fig:pdf} shows the density PDF of warm-hot gas ($10^5 \, {\rm K} \leq T<10^7 \, {\rm K}$) at redshift $z=0$.  At $n_{\rm H}<10^{-4} \, {\rm cm^{-3}}$ (WHIM) the PDF exhibits mild variations among different cosmic structures. However, at higher densities (WCGM) the differences of the PDF among different cosmic structures become negligible. At density $n_{\rm H}>0.13 \, {\rm cm^{-3}}$ the PDF declines because almost all the gas at those densities is star-forming. 

The right panel of Figure~\ref{fig:pdf} shows the density PDF of hot gas $T\geq 10^7 \, {\rm K}$. Our results show that it is strongly peaked in filaments and knots. Since the total mass fraction of hot gas is very small in sheets and voids, the PDFs measured from TNG100 in these regions are not robust. More quantitative conclusions about the PDF of hot gas in sheets and voids could be drawn when cosmological hydrodynamical simulations of much bigger volumes and similar resolution will become available. 

\subsection{Joint Analysis of Cosmic Web Classification and Baryonic Phases}

\begin{figure*}
\begin{center}
 \includegraphics[width=0.48\textwidth]{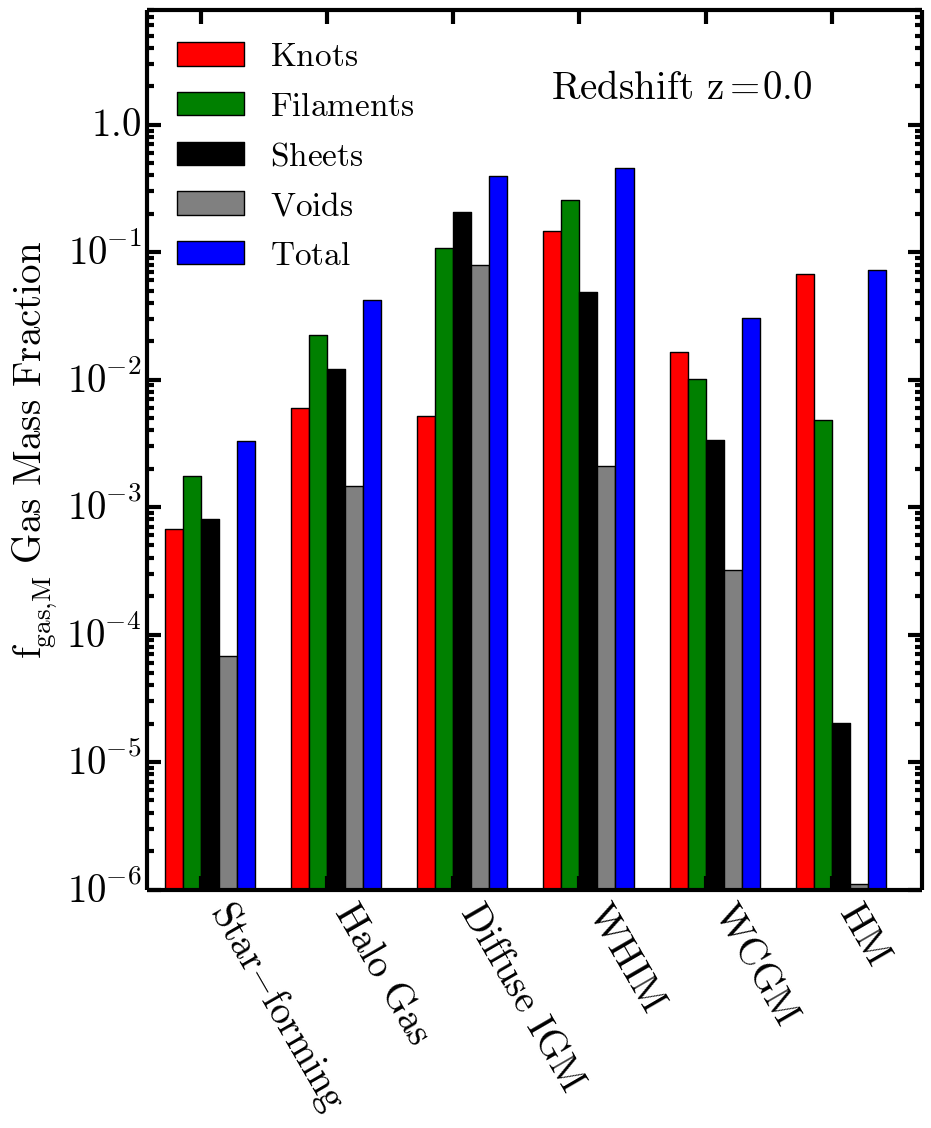}
 \includegraphics[width=0.49\textwidth]{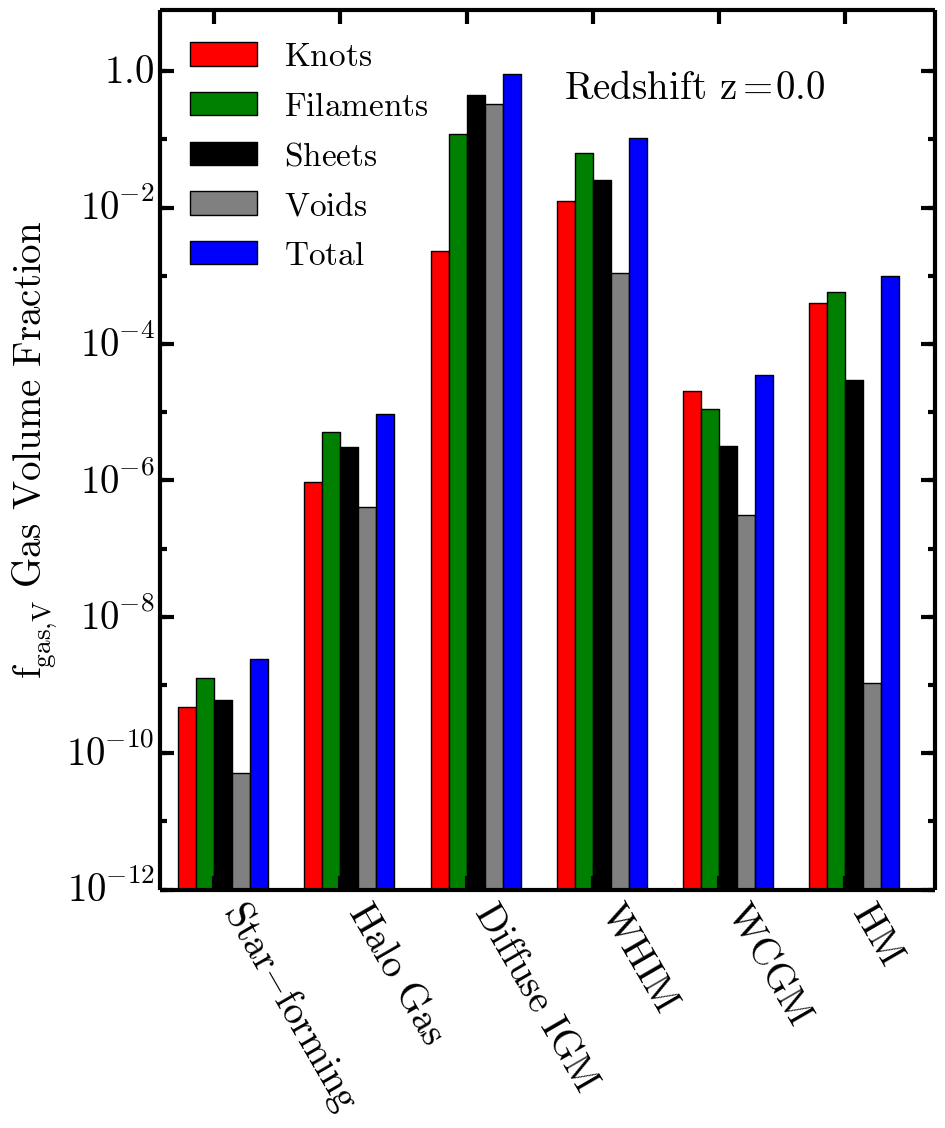}
\end{center}
\caption{ Histograms of the mass and volume fractions of all gaseous baryonic phases in different regions of the Cosmic Web at redshifts $z=0$. Left: mass fractions with respect to the total gas mass budget. Right: volume fractions with respect to the total volume. WHIM and Diffuse IGM dominate the mass and volume baryon budget in the universe. The former dominates in filaments and knots, whereas the latter dominates in sheets and voids.}\label{fig:fractions}
\end{figure*}

More quantitative statements about the state of baryons can be made by measuring the mass/volume fractions associated with each baryonic phase defined in Subsection~\ref{sec:phases} in each region of the Cosmic Web. We show histograms of these quantities in knot, filaments, sheets and voids at redshift $z=0$ in Figure~\ref{fig:fractions}. The mass of Star-forming Gas and Halo Gas peaks in filaments and is a factor $\sim 3.6$ lower in knots. However, the mass of all gas is only $\sim 1.6$ times larger in filaments than in knots, meaning that the difference seen in the Star-forming and Halo gas phases cannot only be attributed to the fact that filaments host more gas. Filaments host more Halo Gas and Star-forming Gas than knots both in absolute and relative terms. The fraction of WCGM peaks in knots and filaments, but is $>5$ times smaller in sheets and voids. HM is mostly found in knots where gas can be successfully shock heated. Figure~\ref{fig:fractions} shows that the WHIM in filaments dominates the mass budget, followed by Diffuse IGM in sheets. The observationally elusive WHIM is present in large fractions in filaments and knots, $\sim 26\%$ and $\sim 15\%$ of the total gas mass, respectively. However, despite the fact that the WHIM hosts more mass than the Diffuse IGM, the former's volume fraction is only $\sim 11\%$, compared to $\sim 89 \%$ for the latter. The WHIM is intrinsically faint and the fact that it occupies a relatively small volume fraction also implies that the probability for quasar sight-lines to intersect WHIM filaments is  lower than the probability of intersecting the Diffuse IGM.

\begin{figure*}
\begin{center}
 \includegraphics[width=0.99\textwidth]{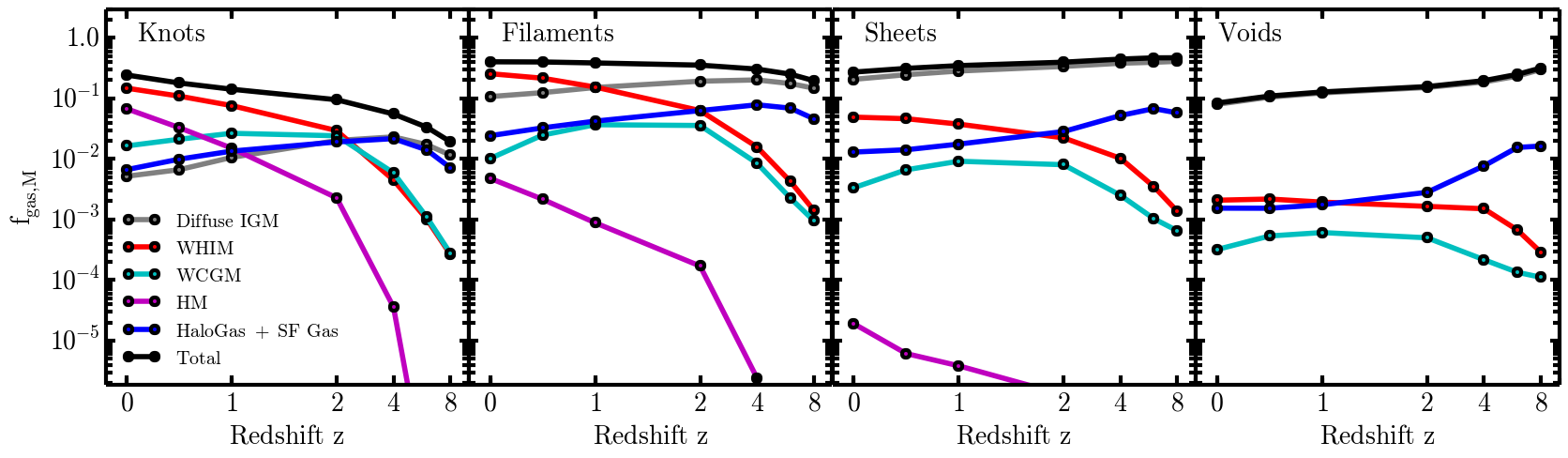}
\end{center}
\caption{ Evolution of the mass fraction of WHIM (red line), Diffuse IGM (gray line), WCGM (cyan line), HM (magenta line) and Halo Gas + Star-forming Gas (blue line) from redshift $z=8$ to redshift $z=0$ in different regions of the Cosmic Web. The evolution of the total gas fraction with respect to the total baryon mass in the universe is shown for for each structure as a black solid line. Panels from left to right: knots, filaments, sheets, voids. The Diffuse IGM and the WHIM dominate the mass budget in all structures at redshift $z=0$. The WHIM is the dominant phase in knots and filaments at $z=0$, but it declines at redshift $z>1$, to be replaced by the Diffuse IGM and the WCGM. The mass fraction of the condensed phase constituted by the Halo Gas and Star-forming Gas is large in all structures at redshift $z>2$, but it relative contribution steadily declines at redshift $z<1$.}\label{fig:fractions_evo}
\end{figure*}

In Figure~\ref{fig:fractions_evo} we show the evolution of the mass fractions of WHIM, Diffuse IGM, WCGM, HM and Halo Gas + Star-forming Gas from redshift $z=8$ to $z=0$. Each panel shows the fraction associated with a different region of the Cosmic Web (knots, filaments, sheets, voids). The mass fraction of the HM is $\gtrsim 20-100$ times smaller than that of WHIM and Diffuse IGM in filaments, sheets and voids at all redshift. However, the abundance of HM is comparable to that of the WHIM in knots at redshift $z\sim 0$, a consequence of the high efficiency of shock heating in these structures. In all regions, the mass fraction of Diffuse IGM declines steadily from high to low redshift. At redshift $z=0$, the mass fraction of WHIM is $\sim 15\%, 26\%, 5\%$ of the total gas mass in the universe in knots, filaments and sheets, respectively. In these cosmic structures the WHIM mass fraction increases by more than a factor of 10 from redshift $z=4$ to redshift $z=0$. In voids, the WHIM mass fraction is $<1\%$ at all epochs and it evolves weakly as a function of redshift. At redshift $1<z<2$ the mass fraction of WCGM peaks, then decreases again. Interestingly, the WCGM fraction in knots and filaments becomes comparable to that of the WHIM at $z>2$, but this result might be susceptible to our choice for the density cut between these two phases at high redshift ($z>2$). The mass fraction of the condensed phases (Halo and Star-forming Gas) constitutes as sub-dominant mass contribution at all redshifts, being comparable to the diffuse IGM mass fraction only in knots, but not in the rest of the cosmic structures. At redshift $z<1$ the condensed phases become sub-dominant with respect to the sum of WHIM and Diffuse IGM in all regions of the Cosmic Web. 

From Figure~\ref{fig:fractions_evo} it is evident that the mass of the Diffuse IGM dominates over that of the WHIM in sheets and voids at all times. However, in filaments and knots, the WHIM becomes the dominant contribution to the cosmic baryon budget at $z<1$. Since the WHIM is expected to represent the bulk of the {\it missing} baryons, it is extremely important to devise future observational campaigns and analysis techniques that are sensitive to signals coming from the shock heated gas in filaments and knots.  

\subsection{Morphology of Baryonic Filaments}

\begin{figure*}
\begin{center}
 \includegraphics[width=0.99\textwidth]{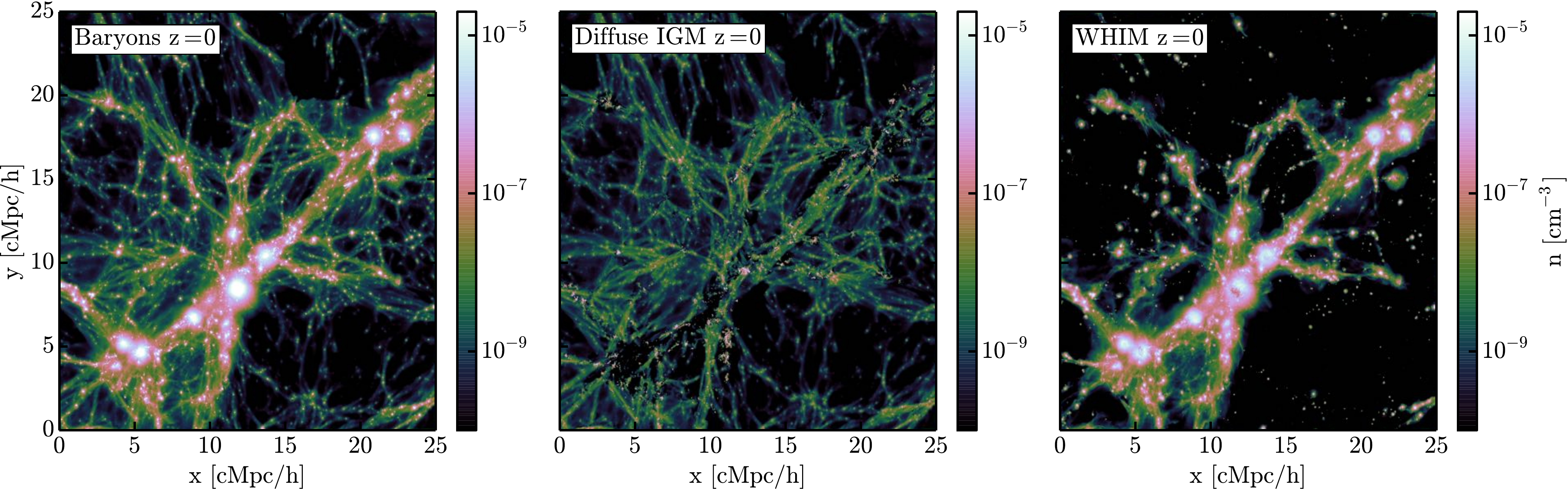}
 \includegraphics[width=0.99\textwidth]{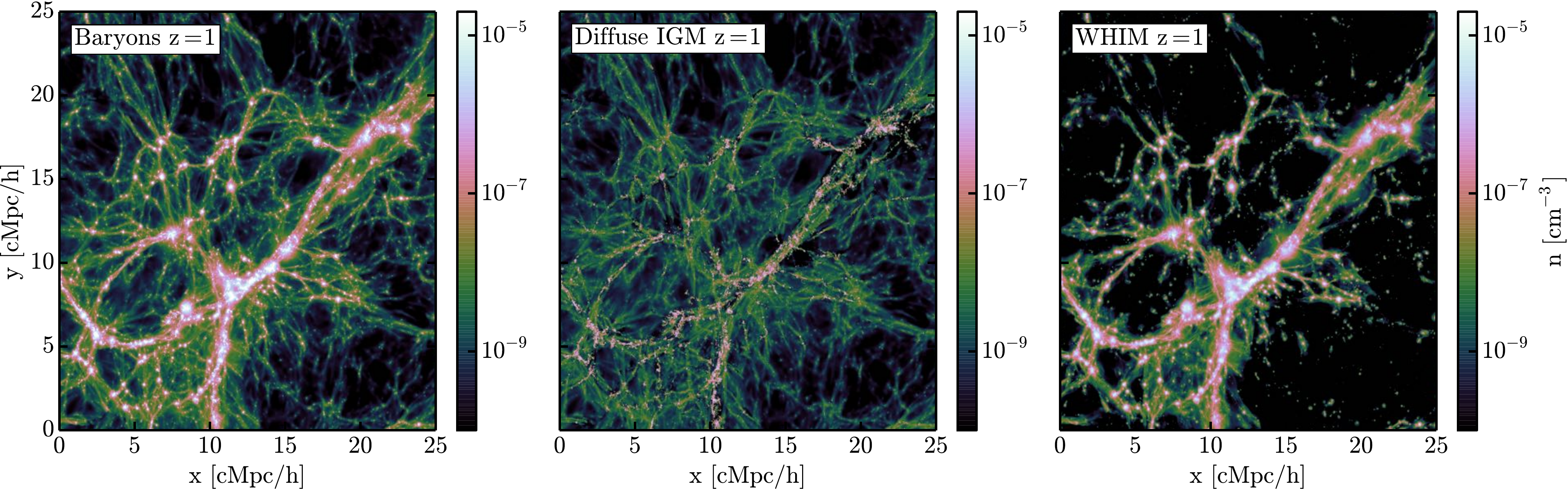}
 \includegraphics[width=0.99\textwidth]{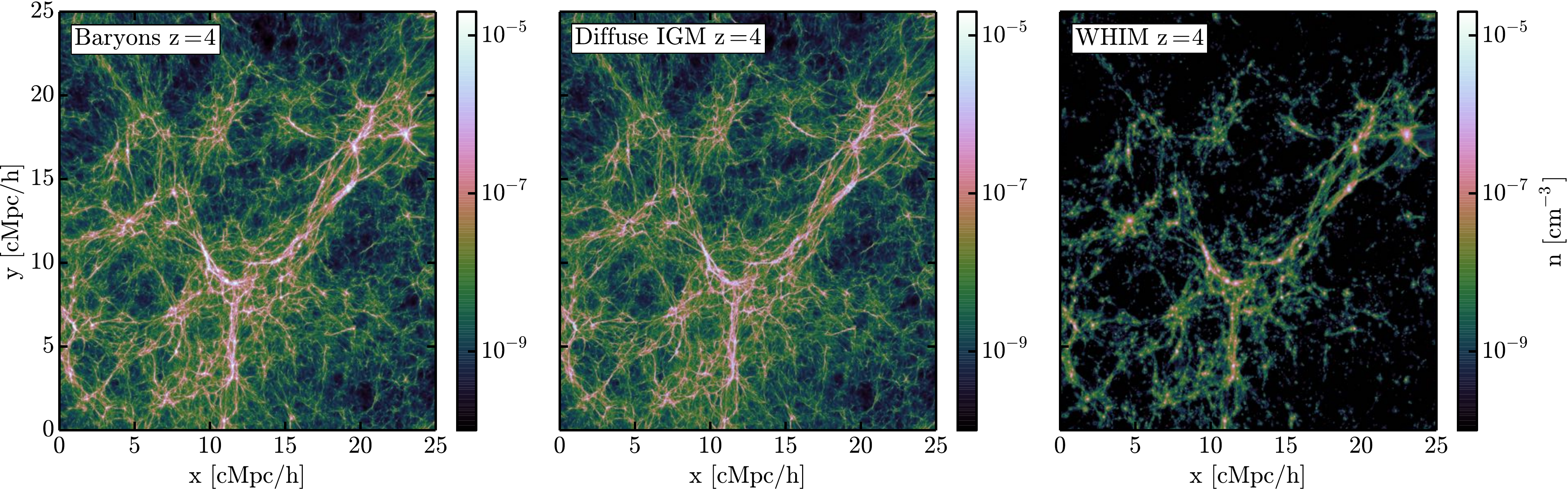}
\end{center}
\caption{Projected number density of a filament in a slice of dimensions $25\times25\times8$ cMpc$^3/h^3$ at redshifts $z=0, 1$, and 4. Left panel: all baryons. Centre: Diffuse IGM. Right: WHIM. By visual inspection we see that the Diffuse IGM is ubiquitous, whereas the WHIM is concentrated near knots and filaments. This result is in agreement with our quantitative measurement of the volume fractions of WHIM and Diffuse IGM (Table~\ref{tab:phases}, Figure~\ref{fig:fractions}).}\label{fig:fil_morph}
\end{figure*}

In the previous Subsections we established that most of the baryonic mass in the universe at redshift $z=0$ is in filaments and that they mostly contain a mixture of WHIM and Diffuse IGM. It is not guaranteed for these two gas phases to be contiguous, especially because the WHIM is expected to be IGM shock heated by accretion onto cosmic structures and by feedback processes. We also showed that the WHIM only occupies a small fraction of the volume compared to the Diffuse IGM. In this Subsection we study the morphology of cosmic filaments. 

Figure~\ref{fig:fil_morph} shows the projected density of gas in a thin slice containing a filament that extends for more than 30 cMpc/$h$ at redshift $z=0$. The same comoving region is also shown at redshifts $z=1$ and 4. The left slices represent the contribution from all baryons. The middle panels show the contribution from the Diffuse IGM. The right panels show the contribution from the WHIM. Comparison between the different density maps highlights how the Diffuse IGM is ubiquitous in the Cosmic Web outside of halos (filaments, sheets, voids). The WHIM is much more confined and is generally found near filaments and around knots, regions where shock heating is more efficiently achieved. These conclusions apply both at redshifts $z=0, 1$ and somewhat more weakly at redshift $z=4$, where the WHIM appears to be sparse and where its mass fraction is small (see previous Subsections). 

\subsection{Metal Enrichment of Extragalactic Baryons}\label{sec:zdf}

\begin{figure*}
\begin{center}
 \includegraphics[width=0.99\textwidth]{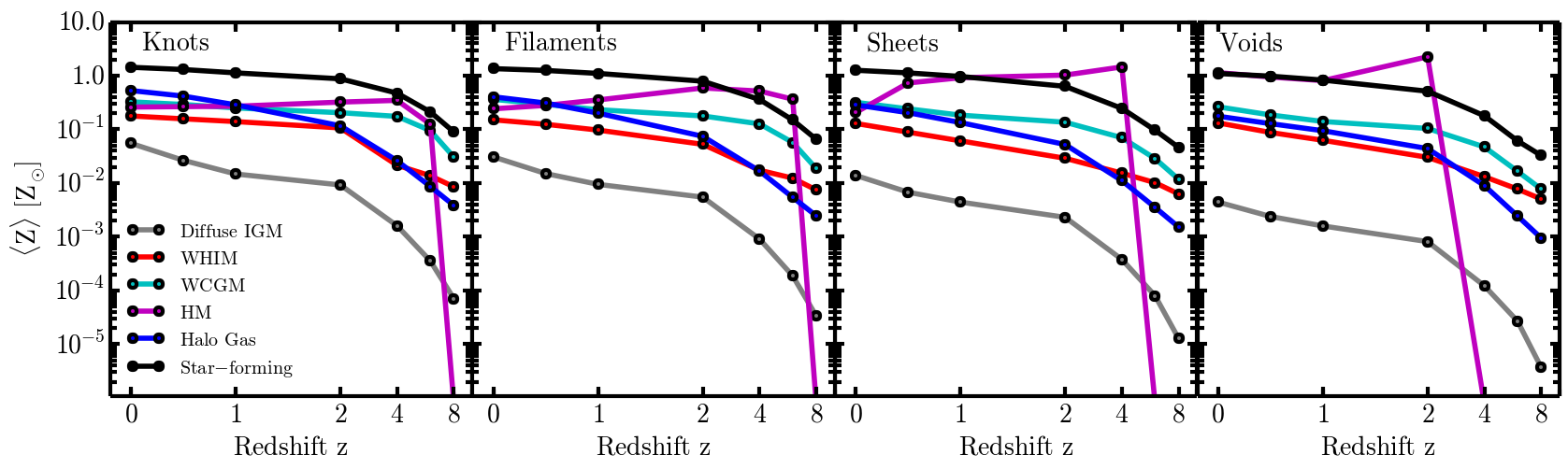}
\end{center}
\caption{ Mean gas metallicity as a function of redshift for knots, filaments, sheets and voids. The red line represents the WHIM, the gray line the Diffuse IGM, the cyan line the WCGM, the magenta line the HM, the blue line the Halo Gas and the black line the Star-forming Gas. The WHIM is more metal rich than the Diffuse IGM. The HM is metal enriched already at redshift $z=4$ and has the highest mean metallicity among all the components at redshift $z\leq 4$. The metallicity of HM in this plot drops sharply at redshift $z>6$ because hardly any HM has formed at this redshift (Figure 7). The Halo Gas and Star-forming Gas have similar degree of metal enrichment and trend with redshift in all regions of the Cosmic Web. Star-forming Gas is more metal rich than the average Halo Gas at all reshifts and in all cosmic structures.}\label{fig:zdf}
\end{figure*}

In this Subsection, we analyse the mean metallicity of the four dominant extragalactic baryonic phases (WHIM, Diffuse IGM, WCGM and HM) in different regions of the Cosmic Web. We choose the mean metallicity as a probe of metal enrichment and its efficiency and give a qualitative interpretation of the results. Figure~\ref{fig:zdf} shows the redshift evolution of the mean metallicity of each baryonic phase in each cosmic structure class. 

The metallicity of the HM is the highest among the different phases considered in Figure~\ref{fig:zdf} and it evolves very weakly at redshift $z\leq 2$ in knots and filaments. In regions where the HM is less abundant (sheets and voids; see previous Subsections) the metallicity decreases from high to low redshift as an effect of dilution with newly shock heated lower metallicity gas. In general, wherever this phase exists, it is already metal enriched at high redshift. This result is in agreement with the completely independent analysis of IllustrisTNG performed by \cite{2018MNRAS.477L..35G} and with work by other authors \citep[e.g.][]{2016MNRAS.459.4408M,2017MNRAS.468..531B, 2018MNRAS.476.2689B}, who also found evidence for early metal enrichment in the progenitors of galaxy cluster galaxies at high redshift.

The metallicity of the WHIM grows monotonically from redshift $z=4$ to redshift $z=0$ in filaments, sheets and voids, but it plateaus at $z\leq 2$ in knots. In filaments and knots at redshift $z=0$ the WHIM is almost as metal rich as the HM. 

The evolution of the metallicity of WCGM is similar to that of the WHIM, but is higher by a factor $\sim 3-5$. This effect is a consequence of its proximity to regions influenced by metal enriched galactic winds \citep{2017arXiv171105261T}.

The metallicity of the Diffuse IGM increases monotonically from redshift $z=8$ to $z=0$ in all structures. This phase is significantly more metal poor than the WHIM. Metal enrichment of the Diffuse IGM is found to be more prominent near knots and filaments. 

Finally, the mean metallicities of the Halo Gas and Star-forming Gas show very similar trends with redshift in all cosmic structures. Interestingly, the mean metallicity of the Star-forming Gas is higher than that of the Halo Gas at all times. This can be interpreted as evidence that on average stars form in regions where gas has been locally enriched by stellar feedback events.

\cite{2017arXiv171105261T} independently analysed the gas metallicity in IllustrisTNG. In their analysis, they defined a warm-hot component with temperature $10^5 \, {\rm K}\leq T \leq 10^7 \, {\rm K}$ (equivalent to our definition of WHIM + WCGM) and a hot phase with temperature $T > 10^7 \, {\rm K}$ (equivalent to our definition of HM). They concluded that the metal mass in the warm-hot phase at redshift $z=0$ is $\sim 4.3$ times larger than the metal mass in the hot phase. Using the values measured in our analysis, the ratio between metal masses measured by \cite{2017arXiv171105261T} can be written as:
\begin{align}
&{\frac{M_{Z,{\rm WHIM}}+M_{Z,{\rm WCGM}}}{M_{Z,{\rm HM}}}\approx} \nonumber \\ 
&{\frac{\langle Z_{\rm WHIM}\rangle f_{\rm WHIM,M}+\langle Z_{\rm WCGM}\rangle f_{\rm WCGM,M}}{\langle Z_{\rm HM}\rangle f_{\rm HM,M}}\sim 4.4}, 
\end{align}
where ${M_{Z,i}}$, ${\langle Z_{i}\rangle }$ and ${f_{i,{\rm M}}}$ are the metal mass, the mass fraction and the mean metallicity of the $i$-th phase, respectively. Reassuringly, we find a value in excellent agreement with that found by \cite{2017arXiv171105261T} despite the use of significantly different post-processing techniques. 

\section{Discussion}\label{sec:discussion}

We compare our work to previous computational work and discuss the observational implications of our results. 

\subsection{Comparison to Previous Computational Work}

Over the last 20 years significant efforts have been spent by the computational cosmology community to characterise the large scale distribution of baryons. We compare to papers that focused explicitly on the cosmic baryon budget. Before going into the details of the comparison, we stress that our WHIM definition is somewhat different from the one used by other authors in the literature. In general, WHIM and WCGM have typically been considered as a common phase by many works cited below.

The seminal papers by \cite{1999ApJ...514....1C} and \cite{1999ApJ...511..521D} used cosmological hydrodynamical simulations to measure the mass and volume fractions of baryons in different phases identified by density and temperature cuts in the phase diagram. Despite the limited number of resolution elements and the simple models for galactic feedback, these simulations offered a clear qualitative picture of the baryon budget at low redshift, dominated by the diffuse IGM and the WHIM. \cite{2001ApJ...552..473D} consolidated these results by performing simulations with spatial resolution $\sim 1-200$ kpc/$h$ with a variety of numerical methods. 

\cite{2006ApJ...650..560C} significantly improved the previous results by performing cosmological hydrodynamical simulations with spatial resolution $\sim 30$ kpc/$h$ and significantly improved models for galactic winds. That paper focuses on the effects of galactic super-winds, concluding that their inclusion increases the abundance of WHIM by $\sim 25\%$ with respect to simulations that do not include them. We find that IllustrisTNG predicts a scenario in qualitative agreement with this work, with the exception of a few minor differences: at redshift $z<1$ IllustrisTNG predicts slightly lower Diffuse IGM mass fraction and slightly higher hot gas fraction as compared to \cite{2006ApJ...650..560C}. This small discrepancy is probably related to the introduction of models for active galactic nuclei feedback, which was not included explicitly by \cite{2006ApJ...650..560C}, and which is responsible for heating gas in massive halos to high temperatures. Furthermore, our analysis allowed us to capture environment-dependent features in the density PDF and metallicity distribution (Subsections \ref{sec:pdf} and \ref{sec:zdf}) that were not captured by the analysis performed by \cite{2006ApJ...650..560C} which was performed by placing simple density and temperature cuts in the baryonic matter phase diagram. 

\cite{2006ApJ...650..560C} and \cite{2007MNRAS.374..427D} analysed the details of the metallicity of baryons in the universe, with comparable results. In particular, \cite{2007MNRAS.374..427D} identified the evolution of the mean metallicity of several baryonic phases at redshift $0 \leq z \leq 6$. We find general agreement with their predictions, even if the effect of environment was not considered in these papers. Our analysis extends their work by demonstrating that the metallicity distribution of cool ($T< 10^5 \, {\rm K}$) and hot ($T> 10^7 \, {\rm K}$) gas depends on the location in the Cosmic Web, whereas the warm-hot phase ($10^5 \, {\rm K} < T < 10^7 \, {\rm K}$) has a metallicity that is nearly independent on the environment. 

Subsequent numerical work with increasingly better resolution and physical modeling of galaxy formation processes has been performed over the years, with more robust results which confirmed and updated the scenario developed in previous work, in particular with respect to the high abundance of WHIM in the universe at redshift $z<1$ \citep{2010MNRAS.408.2051D, 2010MNRAS.402.1911T, 2011ApJ...731....6S, 2012ApJ...759...23S, 2016MNRAS.455.2804S}. 

\begin{figure*}
\begin{center}
 \includegraphics[width=0.99\textwidth]{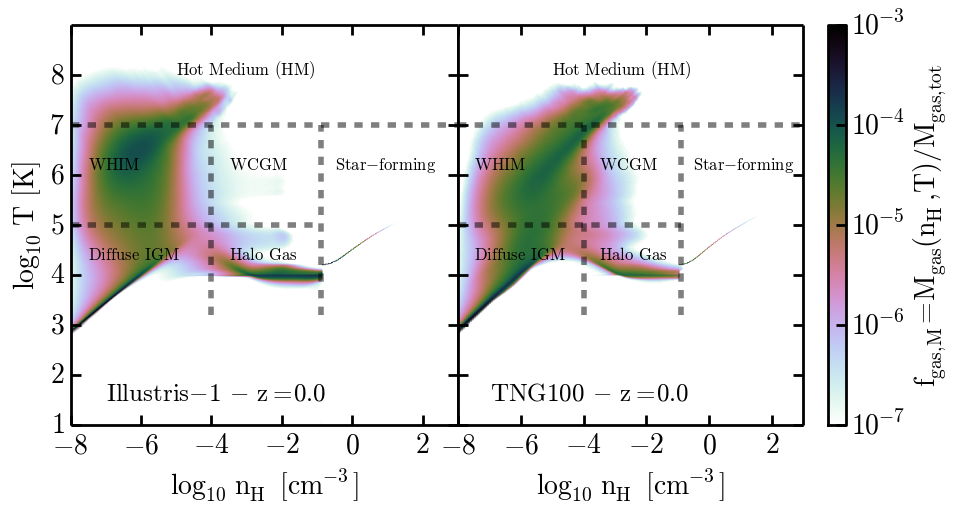}
\end{center}
\caption{ Phase diagram of baryons at redshift $z=0$ for Illustris-1 (left) and TNG100 (right). Baryons in all regions of the Cosmic Web are considered in this plot. The abundance of cool, warm-hot and hot gas varies significantly between the two simulations. }\label{fig:mhist_comp}
\end{figure*}

For our purposes, the most relevant comparison is perhaps the one to the work of \cite{2016MNRAS.457.3024H}, who used the first generation Illustris simulation to study the state of baryons in the Cosmic Web. Given the multiple differences with respect to the analysis of Illustris performed by \cite{2016MNRAS.457.3024H}, we here make a more direct comparison by re-analysing the original Illustris simulation with our pipeline. Figure~\ref{fig:mhist_comp} shows the phase diagram of gas in Illustris and IllustrisTNG at redshift $z=0$. Several differences between the two simulations can be gauged by eye in this plot. In particular, Illustris exhibits (I) a lower abundance of the Diffuse IGM tail, (II) a higher abundance of WHIM, (III) a lower abundance of WCGM, (IV) a higher abundance of condensed gas than IllustrisTNG. By performing a quantitative comparison (Table~\ref{tab:phases-ill-vs-tng}), we find that the mass fraction of hot gas in Illustris is comparable to that of IllustrisTNG. However, the mass in the WHIM in Illustris is $\sim 26\%$ larger than in IllustrisTNG. The mass fraction of the Diffuse IGM is $\sim 43$\% lower in Illustris than in IllustrisTNG. The mass fraction of the WCGM is $\sim 20$ times lower in Illustris compared to IllustrisTNG. Finally, the mass fraction of condensed phases in Illustris is more than twice that in IllustrisTNG. These results indicate that baryons in the IGM are heated to warm-hot temperatures more efficiently in Illustris than in IllustrisTNG, and that the balance between condensed gas in galaxies and WCGM is also offset.

\begin{table}
\centering
\caption{Mass fractions of all gaseous phases of baryonic matter, $f_{\rm gas,M}$ in Illustris and IllustrisTNG. The mass fractions are normalised with respect to the total gas mass. Results are reported for redshift $z=0$ and are achieved by integrating over all regions of the Cosmic Web, so the values do not depend on the classification algorithm. }\label{tab:phases-ill-vs-tng}
{\bfseries Gas Mass Fractions in Illustris vs. IllustrisTNG}
\makebox[\linewidth]{
\begin{tabular}{llcc}
\hline
\hline
\multicolumn{4}{l}{Redshift $z=0.0$} \\ 
\hline
 & Phase & Illustris & IllustrisTNG \\
\hline
 & Star-forming Gas & $9.0\times 10^{-3}$ & $3.3\times 10^{-3}$ \\
 & Halo Gas & $1.1\times 10^{-1}$ & $4.3\times 10^{-2}$ \\
 & Diffuse IGM & $2.2\times 10^{-1}$ & $3.9\times 10^{-1}$ \\
 & WHIM & $5.8\times 10^{-1}$ & $4.6\times 10^{-1}$ \\
 & WCGM  & $1.6\times 10^{-3}$ & $3.1\times 10^{-2}$ \\
 & HM  & $7.3\times 10^{-2}$ & $7.3\times 10^{-2}$ \\
\hline
\hline
\end{tabular}
}
\end{table}

The original Illustris simulation had particularly strong active galactic nucleus feedback, that was also responsible for heating and ejecting large amounts of baryons from massive galaxy clusters, producing cluster baryon fractions lower than in observations \citep{2014MNRAS.445..175G}. The improvements in the feedback schemes introduced in IllustrisTNG significantly alleviated this problem \citep{2017MNRAS.465.3291W, 2018MNRAS.473.4077P}. As a result, extragalactic gas in IllustrisTNG is heated less efficiently by feedback processes than in the original Illustris. For this reason, we attribute the differences in the Diffuse IGM and WHIM mass fraction between Illustris and IllustrisTNG to the different efficiency of black hole feedback. These results may indicate that high-temperature and/or x-ray probes of hot gas in-and-around galaxies may give important insight into the nature of active galactic nucleus feedback. Additionally, the prescriptions for galactic winds were also updated significantly between Illustris and IllustrisTNG, a possible origin of the differences seen in the dense gas in galaxies and their local environment (condensed phases and WCGM). A detailed comparison of models with different sub-resolution physics is beyond the scope of this paper, and we leave it for future work.

\cite{2018MNRAS.473.4077P} compared the outcomes of the Illustris and IllustrisTNG models to a broad range of observations (galaxy mass function, star formation density, gas fractions in dark matter halos), concluding that IllustrisTNG provides a better match to observational data. This comparison highlights how the improvements in the feedback models introduced in IllustrisTNG were crucial to produce a more realistic simulated galaxy population and gas distribution. For these reasons, we believe IllustrisTNG provides more reliable predictions for these baryonic phases.

\subsection{Observational Implications for the Hunt for the WHIM}

\begin{figure*}
\begin{center}
 \includegraphics[width=0.49\textwidth]{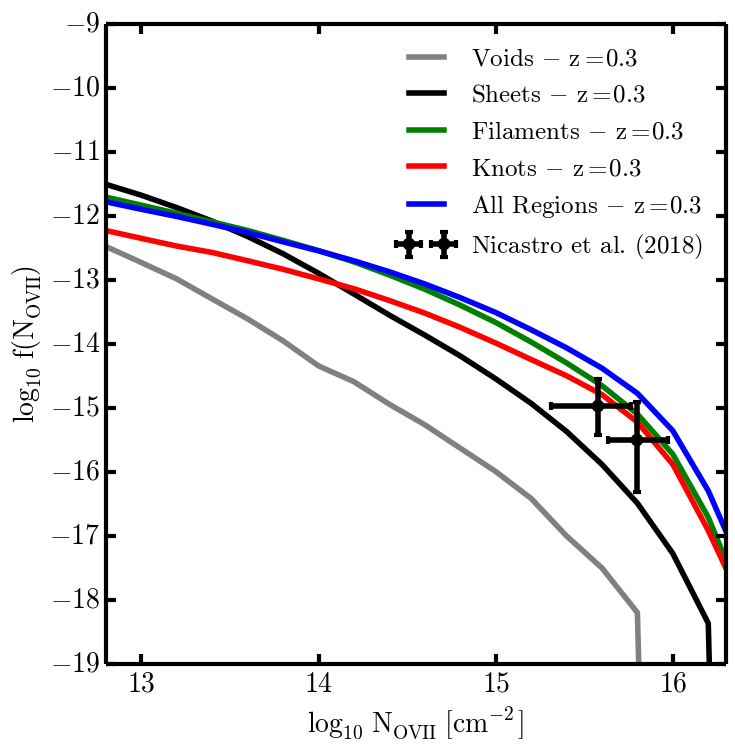}
 \includegraphics[width=0.49\textwidth]{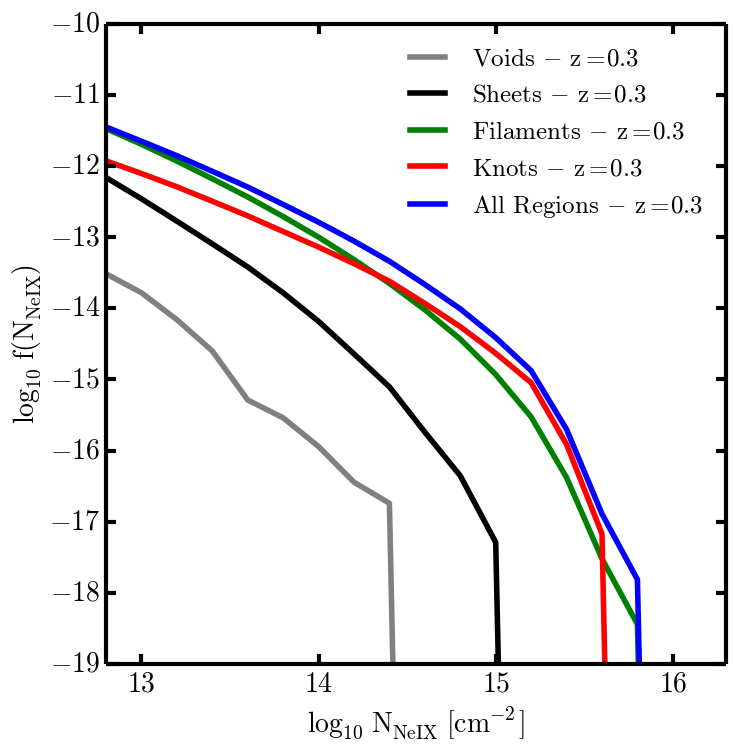}
\end{center}
\caption{ The column density distribution function (CCDF) of OVII (left panel) and NeIX (right panel) at redshift $z=0.3$ from gas in different regions of the Cosmic Web. Nelson et al. (2018a) showed similar plots for the oxygen ions from the IllustrisTNG simulations, but did not include information about the location in the Cosmic Web. Column densities on the x-axis are computed only considering gas in each class of cosmic structures, therefore the CCDFs are not additive. Column densities are computed by integrating over the whole depth of the simulation, so they are only  upper limits to the actual column densities from multiple absorbers; for this reason the CDDFs should only be used to appreciate differences between absorbers in different regions of the Cosmic Web. Two WHIM OVII absorbers at redshift $z\sim 0.3-0.4$ have been detected for the first time by Nicastro et al. (2018); their data are reported in the left panel. NeIX absorbers are currently not detected in the IGM, but we predict that they may be found in filaments and sheets, with a significant contribution in knots only at high column densities. }\label{fig:cddf}
\end{figure*}

Roughly half of the baryons at redshift $z<1$ have not been explicitly detected \citep{1998ApJ...503..518F, 2004IAUS..220..227F, 2007ARA&A..45..221B} and for more than a decade cosmological simulations have been suggesting that they may be found in the WHIM. For this reason, 
significant effort has been spent in developing methods to measure the abundance of intergalactic baryons with alternative techniques. 

One of the most widely used techniques to identify intergalactic baryons is to look for absorption lines in the spectra of background sources like quasars. Looking for absorption lines from specific ions allows one to probe intergalactic plasma in different phases and characterise its properties. Over the years, significant success has been achieved in the detection of hydrogen and metal absorbers along quasar lines of sight. The column density of neutral hydrogen in these absorbers can be constrained by looking for Ly-$\alpha$ absorption features, whereas warmer diffuse gas can be traced with OVI UV absorption \citep{2014ApJ...794...75S, 2015ApJ...806...25S, 2016ApJ...818..113N, 2016ApJ...817..111D}. Attempts have been made to probe the WHIM using X-ray spectroscopy aimed at detecting absorption lines from oxygen's higher ionization states OVII and OVIII, but results have been largely inconclusive \citep{2001ApJ...554L...9P, 2003PASJ...55..879Y, 2005MNRAS.360.1110V, 2009ApJ...697..328B, 2011ApJ...734...91T, 2013arXiv1306.2324K}. However, \cite{2018Natur.558..406N} recently reported the detection of two OVII absorbers at redshift $z\sim 0.3-0.4$ which were found to be consistent with WHIM absorber after ionization modeling. With the advent of the next generation X-ray telescopes like Athena, it will be possible to better detect WHIM absorbers from highly ionized species (e.g. oxygen, neon).
 
\cite{2018MNRAS.477..450N} analysed in great detail the spatial distribution and statistics of ionized oxygen (OVI, OVII, OVIII). Here, we focus only on OVII and NeIX that trace the WHIM. We use the ion densities estimated with the method of Subsection~\ref{sec:ions} to compute the column density distribution function (CDDF) of OVII and NeIX absorbers at redshift $z=0.3$. The $512^3$ cubic grids containing 3-d ion densities from Subsection~\ref{sec:ions} are used to compute column densities. For  three faces of the cube orthogonal to the x, y and z axes, respectively, we compute $512^2$ column densities by integrating each 3-d ion density along the axis orthogonal to the face. As a result, we obtain $3\times 512^2$ total columns for each ion. Notice that we do not separate multiple absorbers along the same column in velocity space as in \cite{2018MNRAS.477..450N}, so the measured column densities are only upper limits if a specific line of sight intersects more than one absorber. Since the 3-d cubic ion density grids have the same spacing as those used for the Cosmic Web classification, we can also  integrate along each line of sight by selectively including only the gas from a specific type of cosmic structure. Following this approach, for each line of sight we also compute column densities obtained by only including gas in knots, filaments, sheets and voids, respectively. After the column densities have been computed, we compute the CDDFs following \cite{2018MNRAS.477..450N}. Notice that for each ion the CDDFs of gas from different regions of the Cosmic Web do not add up to give the total CDDF, because only column densities can be added, but not their distributions. Our CDDFs should not be used for detailed comparison to data, because of the simplified approach used to compute column densities, but they can be used to better understand the individual contributions from different cosmic structures. The results of this analysis are shown in Figure~\ref{fig:cddf}. 

Figure~\ref{fig:cddf} shows the abundance of OVII and NeIX, which can in principle be detected with future X-ray instruments. OVII is mainly found in filaments and sheets, with a significant contribution in knots at ${N_{\rm OVII}>10^{15} \, {\rm cm}^{-2}}$. We compare the results from IllustrisTNG to the recent measurements of OVII absorbers in quasar spectra at redshift $z\sim 0.3-0.4$ from \cite{2018Natur.558..406N}. In their paper, these authors do not report the CDDF but the cumulative distribution of OVII absorbers per unit redshift with equivalent width larger than ${\rm EW}$, $dn_{\rm OVII}(>{\rm EW})/dz$. To convert this information to the points with error bars shown in Figure~\ref{fig:cddf}, we made a number of assumptions. First of all, we use the absorber oxygen column densities $N_{\rm O}$, the OVII ionization fraction $f_{\rm OVII}$ inferred by \cite{2018Natur.558..406N} via ionization modeling and quoted in their text. The OVII column densities are then calculated as $N_{\rm OVII}=f_{\rm OVII}N_{\rm O}$. Since we expect the CDDF to quickly decline at $N_{\rm OVII}>10^{16} \, {\rm cm^{-3}}$, the cumulative number of absorbers of \cite{2018Natur.558..406N} is approximately equal to the differential number count in a logarithmic column density bin near the column density of the absorbers. Under this assumption, the OVII CDDF from \cite{2018Natur.558..406N} is estimated as: 
$$
f_{\rm data}(N_{\rm OVII})\approx \frac{dn_{\rm OVII}(>{\rm EW})}{dz\Delta N_{\rm OVII}}\approx\frac{dn_{\rm OVII}(>{\rm EW})}{dzN_{\rm OVII}\Delta\log N_{\rm OVII}},
$$
where the logarithmic bin is chosen to be $\Delta\log N_{\rm OVII}=1$. Taking into account the approximate conversion performed on the observational data, we conclude that the total OVII CDDF from IllustrisTNG compares well to the data from \cite{2018Natur.558..406N}. 

Figure~\ref{fig:cddf} shows that NeIX absorbers have column densities smaller than OVII, ${N_{\rm NeIX}<10^{16} \, {\rm cm}^{-2}}$ and can be found in similar ratios in knots and filaments. Notice that the CDDFs of OVII and NeIX in each type of cosmic structure have different shapes, implying that the two ions trace the Cosmic Web differently. In particular, the OVII CDDF in filaments is larger than the OVII CDDF in other structures at all column densities, whereas the NeIX CDDF in filaments is lower than the NeIX CDDF in knots at ${N_{\rm NeIX}>10^{14.5} \, {\rm cm}^{-2}}$. We will explore the implications in more detail in future work.

\section{Summary and conclusions}\label{sec:summary}

In this paper we performed a joint analysis of the phase of baryons in different classes of structures in the Cosmic Web. The analysis was performed on the TNG100 cosmological, hydrodynamical simulations which follows the evolution of dark matter and baryons in large cosmological volumes. We developed a classification scheme based on the deformation tensor \citep{2007MNRAS.375..489H, 2009MNRAS.396.1815F} which reliably identifies knots, filaments, sheets and voids in the Cosmic Web. We characterised the state of baryons in different phases in each class of cosmic structure. Our work is a significant update and generalization of previous analysis \citep{1999ApJ...514....1C, 1999ApJ...511..521D, 2001ApJ...559L...5C, 2001ApJ...552..473D, 2005ApJ...620...21K, 2006ApJ...650..560C, 2007MNRAS.374..427D, 2010MNRAS.408.2051D, 2010MNRAS.402.1911T, 2011ApJ...731....6S, 2012ApJ...759...23S, 2016MNRAS.462..448G}. 

The smallest contribution to the total baryon budget at redshift $z<1$ comes from the Halo Gas and Star-forming Gas inside and in proximity of galaxies. Interestingly, we find that filaments host more Halo Gas and Star-forming Gas than knots, and they also have a higher relative mass fraction of gas in these phases than the other cosmic structures. However, since the Halo Gas and the Star-forming Gas are highly condensed and our analysis focused on the large scale gas distribution, our quantitative results for these phases may be susceptible to the choice of the size of Gaussian kernel used to smooth the density field and classify the Cosmic Web.

Our results confirm the previous findings that the cosmic baryon budget at redshift $z\leq 1$ is largely dominated by the IGM, in particular its cool phase ($T<10^5$ K) and  warm-hot phase (WHIM, $10^5 \, {\rm K} <T<10^7 \, {\rm K}$, $n_{\rm H}<10^{-4}(1+z)\, {\rm cm}^{-3}$). We demonstrate that the former is much more volume filling than the latter. The WHIM is shock-heated material that mostly resides in filaments and knots and constitutes $\sim 46\%$ of all the baryons in the universe at $z=0$, but it becomes sub-dominant with respect to cooler IGM at $z>1$. 

The warm-hot dense gas ($10^5 \, {\rm K} <T<10^7\, {\rm K}$, $n_{\rm H} >10^{-4} (1+z)\, {\rm cm^{-3}}$) represents only a few percent of all the baryons, it is found mostly in filaments and knots, and it is significantly more metal enriched than the WHIM, which is probably related to the fact that the WCGM might contain larger reservoirs of metal enriched material from galactic winds. This hypothesis needs to be tested with a more detailed analysis. 

We found that most of the hot gas ($T>10^7 \, {\rm K}$, $\sim 7.3\%$ of all the gaseous baryons) in the low redshift universe is found in the knots of the Cosmic Web and that this phase becomes highly sub-dominant at high redshift, where large halos that can shock-heat the baryons to temperatures $T>10^7 \, {\rm K}$ are rare. Nonetheless, small amounts of hot gas are available in the knots of the Cosmic Web at redshift $z=4$ where we also found evidence for early metal enrichment. The hot gas is the component whose density distribution varies the most between knots, filaments, sheets and voids, a feature that may be related to the abundance of massive galaxy clusters and their hot ICM in these structures, and to the presence of active galactic nuclei that can also heat significant amounts gas. 

Finally, we examined the distribution of column densities of OVII and NeIX which trace the WHIM. OVII column density distributions in IllustrisTNG were already analysed by \cite{2018MNRAS.475..624N}, but we performed a complementary analysis by including information on the location of absorbers in the Cosmic Web. Intergalactic OVII is typically found in filaments. NeIX can be observed in knots and filaments in similar ratios. This prediction is particularly relevant to design observational campaigns to detect OVII and NeIX which trace the WHIM and should be detectable via X-ray absorption spectroscopy with future instruments like the Athena X-ray Observatory. 

Our results motivate future research in multiple directions. First of all, our work constitutes a significant update of the theoretical picture that motivates the search for the {\it missing} baryons in the low redshift universe. In future papers, we will focus on designing new techniques to measure and characterise the baryons in cosmic filaments and therefore directly detect half of the baryons in the low redshift universe (the WHIM), an effort that has only recently started to give its fruits \citep{2018MNRAS.475.3854G, 2017arXiv170910378D}. Finally, 
our work can be extended to include galaxies and their properties as they relate to the Cosmic Web \citep[see e.g.][]{2018MNRAS.474..547K}. In fact, galaxies and feedback processes triggered within them are expected to significantly influence the properties of baryons in their environment, in particular their metallicities. Analysis of the metal enrichment process in conjunction with the properties of galaxies and their location in the Cosmic Web will provide a significantly updated theoretical picture of cosmic structure evolution, and it is within reach with the tools we developed for the present work.

%%%
\section*{Acknowledgments}
We thank the editors and the reviewer for their useful comments and suggestions, which contributed to improve the quality of this paper. DM was supported by the CTA and DARK-Carlsberg Foundation Fellowship. DM acknowledges contribution from the Danish council for independent research under the project ``Fundamentals of Dark Matter Structures'', DFF - 6108-00470. MCA acknowledges financial support from the Austrian National Science Foundation through FWF stand-alone grant P31154-N27. PT acknowledges support from NASA through Hubble Fellowship grant HST-HF2-51341.001-A awarded by STScI, which is operated under contract NAS5-26555. RW and VS acknowledges support through the European Research Council under ERCStG grant EXAGAL-308037, and would like to thank the Klaus Tschira Foundation. The IllustrisTNG simulations and the ancillary runs were run on the HazelHen Cray XC40-system (project GCS-ILLU), Stampede supercomputer at TACC/XSEDE (allocation AST140063), at the Hydra, Draco supercomputers at the Max Planck Computing and Data Facility, and on the MIT/Harvard computing facilities supported by FAS and MIT MKI.
%%%
\bibliography{main}

\appendix
%%%

\section{Temperature correction for the low density gas}\label{appendix:tcorr}

As we discussed in Section~\ref{sec:phases}, the results of this paper use the updated values for the thermal energy of the fluid, that correct for a small numerical error in the IllustrisTNG simulations that influences the temperature of the low density gas. The original and updated values are both provided by the IllustrisTNG public data release (\citealt[][for details on the correction]{2018arXiv181205609N}). The correction uses results from a TNG model variant box of size 25 Mpc/$h$ and $512^3$ resolution. This simulation was run with a version of AREPO that includes the fix for this issue. The adiabat of the low density gas was then identified in all TNG runs as well as in the corrected simulation. A multiplicative correction $f_{\rm corr}=T_{\rm corr}/T_{\rm orig}$, the ratio between the corrected linear gas temperature $T_{\rm corr}$ and uncorrected linear gas temperature $T_{\rm orig}$, is then defined and applied as a function of density. The correction is only applied to gas with physical hydrogen number density $n_{\rm H}< 10^{-6}(1+z)^{3} \, {\rm cm}^{-3}$, where the redshift-dependent factor has been introduced to take cosmological expansion into account. The correction is further restricted to the low-temperature IGM by applying the following smooth damping of $f_{\rm corr}$ to unity by $T_{\rm orig}\approx 10^5 \, {\rm K}$:
\begin{equation}
\log T_{\rm corr} = \log T_{\rm orig} + f_{\rm corr}w(T_{\rm orig}),  
\end{equation}
where
\begin{equation}
w(T_{\rm orig})=1-\left\{ \tanh [5(\log T_{\rm orig}-5)] +1 \right\} /2.
\end{equation}

\begin{figure*}
\begin{center}
 \includegraphics[width=0.8\textwidth]{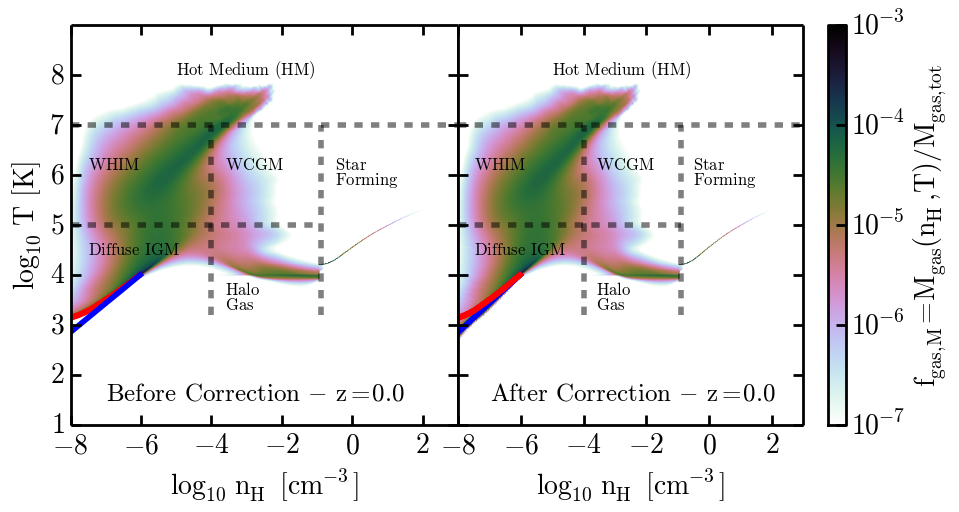}
 \includegraphics[width=0.8\textwidth]{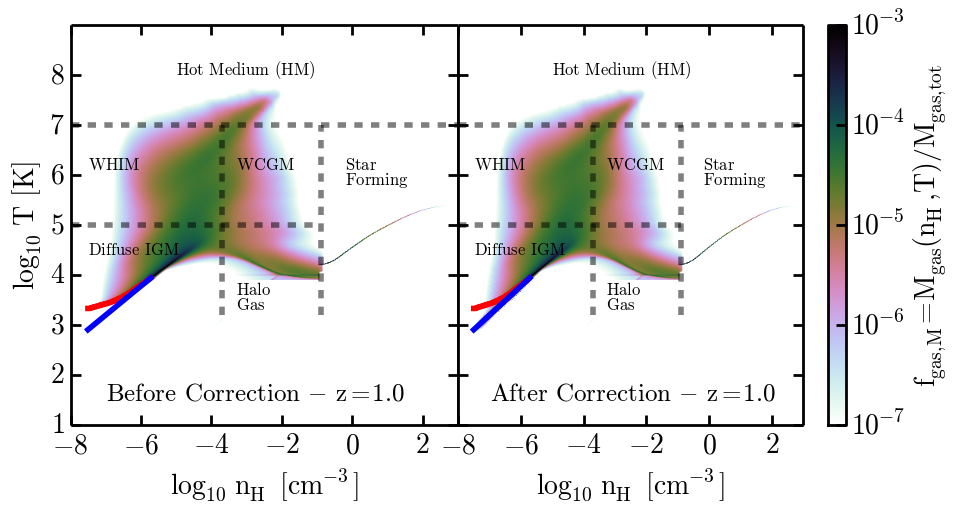}
 \includegraphics[width=0.8\textwidth]{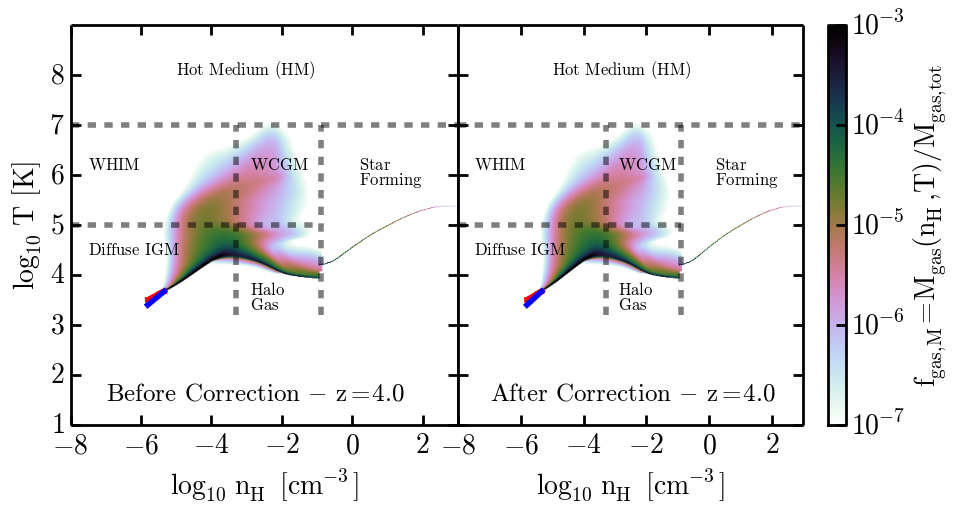}
\end{center}
\caption{ Phase diagram of baryons at redshift $z=0, 1$, and 4 in different regions before (left panels) and after the temperature correction (right panels). The red line represents the original IGM adiabat, whereas the blue line represents the IGM adiabat after correction. The magnitude of the correction appears to decrease at redshift $z\geq 1$.}\label{fig:phase-tcorr}
\end{figure*}

The magnitude of $f_{\rm corr}$ at multiple redshifts can be appreciated in Figure~\ref{fig:phase-tcorr}. The plot shows that the general properties of the phase diagram are preserved, and that only gas near the adiabat of the Diffuse IGM is influenced. 

Most of the quantitative results in this paper are based on the gas mass fraction of each phase. In order to assess the impact of the temperature correction on our results, we performed a direct comparison of the mass fractions for each gas phase before and after the temperature correction (Table~\ref{tab:phases-corr}). We find that the differences are negligible, because we focus on integrated quantities, rather than the detailed structure of the phase diagram. 

%%%%%%%%%%%%%%%%%%%%%%

\begin{table}
\centering
\caption{Mass fractions of all gaseous phases of baryonic matter, $f_{\rm gas,M}$ in TNG100 before the IGM temperature correction (left) and after the temperature correction (right). The mass fractions are normalised with respect to the total gas mass. Results are reported for redshift $z=0, 1, 2, 4$ and are achieved by integrating over all regions of the Cosmic Web, so the values do not depend on the classification algorithm. }\label{tab:phases-corr}
{\bfseries Gas Mass Fractions in TNG100 \\ Effect of IGM Temperature Correction}
\makebox[\linewidth]{
\begin{tabular}{llcc}
\hline
\hline
\multicolumn{4}{l}{Redshift $z=0.0$} \\ 
\hline
 & Phase & Before Correction & After Correction \\
\hline
 & Star-forming Gas & $3.3\times 10^{-3}$ & $3.3\times 10^{-3}$ \\
 & Halo Gas & $4.3\times 10^{-2}$ & $4.3\times 10^{-2}$ \\
 & Diffuse IGM & $3.8\times 10^{-1}$ & $3.9\times 10^{-1}$ \\
 & WHIM & $4.7\times 10^{-1}$ & $4.6\times 10^{-1}$ \\
 & WCGM  & $3.1\times 10^{-2}$ & $3.1\times 10^{-2}$ \\
 & HM  & $7.3\times 10^{-2}$ & $7.3\times 10^{-2}$ \\
\hline
\hline
\multicolumn{4}{l}{Redshift $z=1.0$} \\ 
\hline
 & Phase & Before Correction & After Correction \\
\hline
 & Star-forming Gas & $1.0\times 10^{-2}$ & $1.0\times 10^{-2}$ \\
 & Halo Gas & $6.3\times 10^{-2}$ & $6.3\times 10^{-2}$ \\
 & Diffuse IGM & $5.6\times 10^{-1}$ & $5.8\times 10^{-1}$ \\
 & WHIM & $2.8\times 10^{-1}$ & $2.6\times 10^{-1}$ \\
 & WCGM  & $7.2\times 10^{-2}$ & $7.2\times 10^{-2}$ \\
 & HM  & $1.6\times 10^{-2}$ & $1.6\times 10^{-2}$ \\
\hline
\hline
\multicolumn{4}{l}{Redshift $z=2.0$} \\ 
\hline
 & Phase & Before Correction & After Correction \\
\hline
 & Star-forming Gas & $1.3\times 10^{-2}$ & $1.3\times 10^{-2}$ \\
 & Halo Gas & $9.6\times 10^{-2}$ & $9.6\times 10^{-2}$ \\
 & Diffuse IGM & $7.0\times 10^{-1}$ & $7.1\times 10^{-1}$ \\
 & WHIM & $1.2\times 10^{-1}$ & $1.1\times 10^{-1}$ \\
 & WCGM  & $6.5\times 10^{-2}$ & $6.5\times 10^{-2}$ \\
 & HM  & $2.3\times 10^{-3}$ & $2.3\times 10^{-3}$ \\
\hline
\hline
\multicolumn{4}{l}{Redshift $z=4.0$} \\ 
\hline
 & Phase & Before Correction & After Correction \\
\hline
 & Star-forming Gas & $8.0\times 10^{-3}$ & $8.0\times 10^{-3}$ \\
 & Halo Gas & $1.5\times 10^{-1}$ & $1.5\times 10^{-1}$ \\
 & Diffuse IGM & $7.9\times 10^{-1}$ & $7.9\times 10^{-1}$ \\
 & WHIM & $3.2\times 10^{-2}$ & $3.2\times 10^{-2}$ \\
 & WCGM  & $1.7\times 10^{-2}$ & $1.7\times 10^{-2}$ \\
 & HM  & $3.8\times 10^{-5}$ & $3.8\times 10^{-5}$ \\ 
\hline
\hline
\end{tabular}
}
\end{table}

%%%

\section{Tests of the Cosmic Web classifier}\label{appendix:cwebtests}

\begin{figure*}
\begin{center}
 \includegraphics[width=0.99\textwidth]{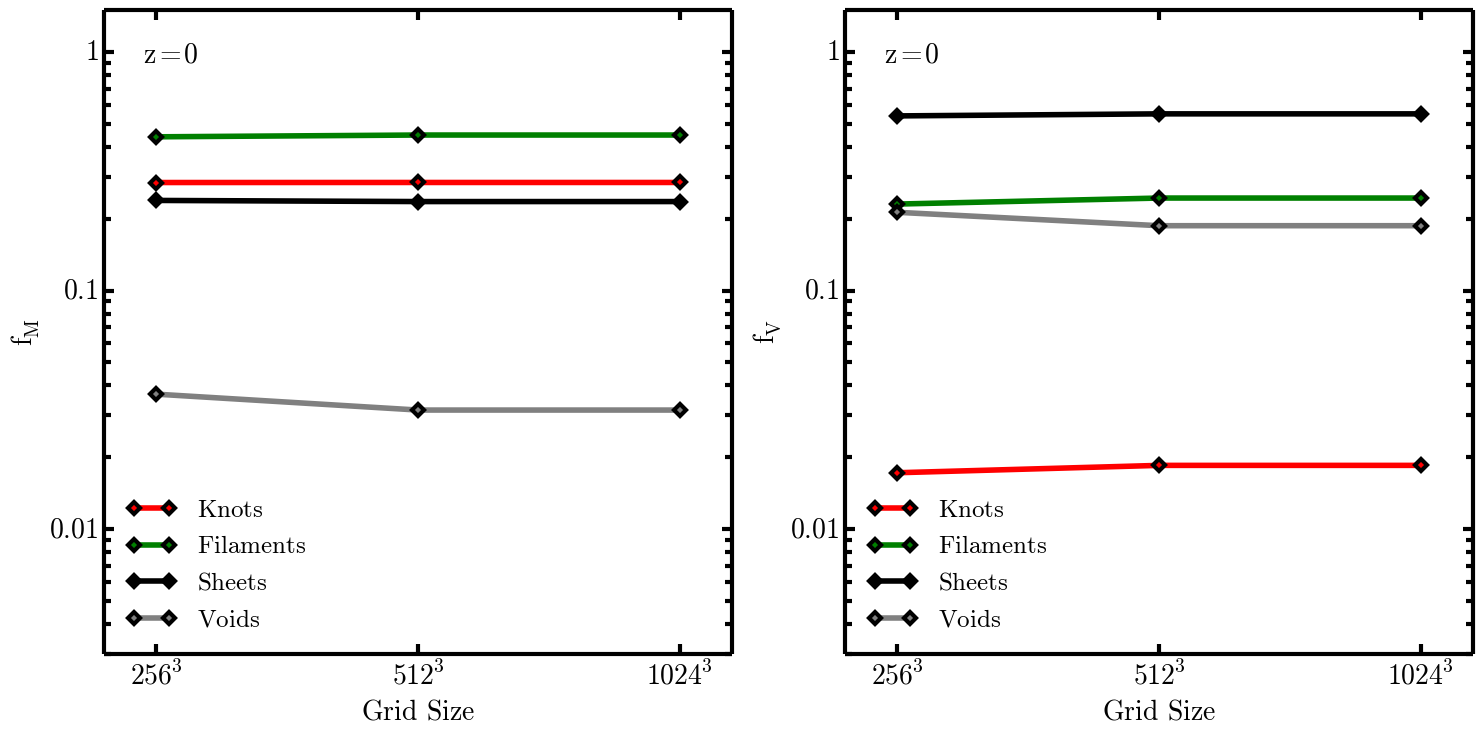}
\end{center}
\caption{ Mass (left panel) and volume (right panel) fractions of voids, sheets, filaments and knots for different Cartesian grid sizes for the classification algorithm. Cosmic Web classification has been performed with eigenvalue threshold $\lambda_{\rm th}=0.07$ and smoothing scale $R_{\rm G}=4 \, {\rm cMpc}/h$ for this plot.}\label{fig:conv_resolution}
\end{figure*}

\begin{figure*}
\begin{center}
 \includegraphics[width=0.99\textwidth]{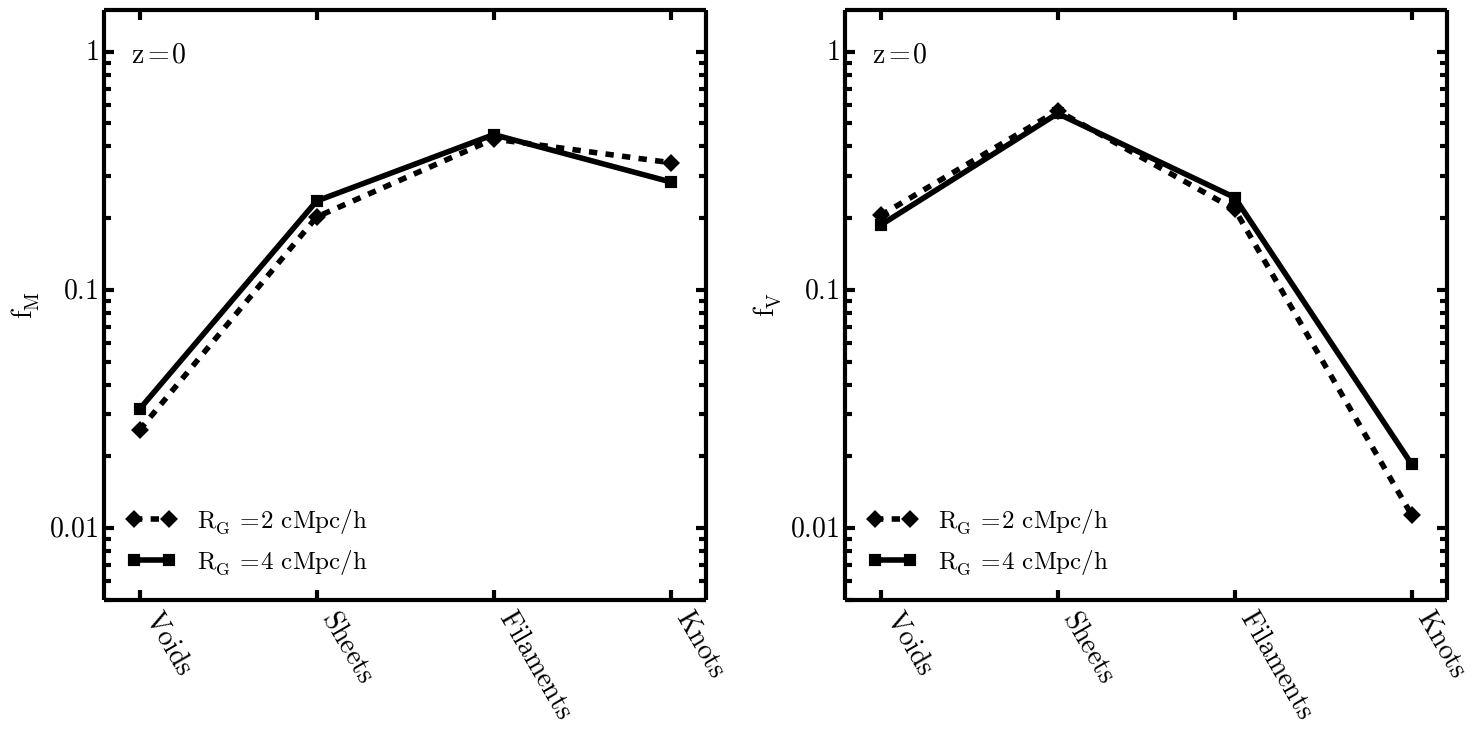}
\end{center}
\caption{ Mass (left panel) and volume (right panel) fractions of voids, sheets, filaments and knots for different smoothing lengths for the classification algorithm. Cosmic Web classification has been performed with eigenvalue threshold $\lambda_{\rm th}=0.07$ and grid size $512^3$ for this plot.}\label{fig:conv_smoothing}
\end{figure*}

We performed several tests to assess the accuracy and robustness of the method we adopted to classify the Cosmic Web. In this Appendix, we show the results of our tests that focus on the mass and volume fractions of different cosmic structures, ${\rm f_{M}}$ and ${\rm f_{V}}$, respectively, at redshift $z=0$. 

\begin{figure*}
\begin{center}
 \includegraphics[width=0.99\textwidth]{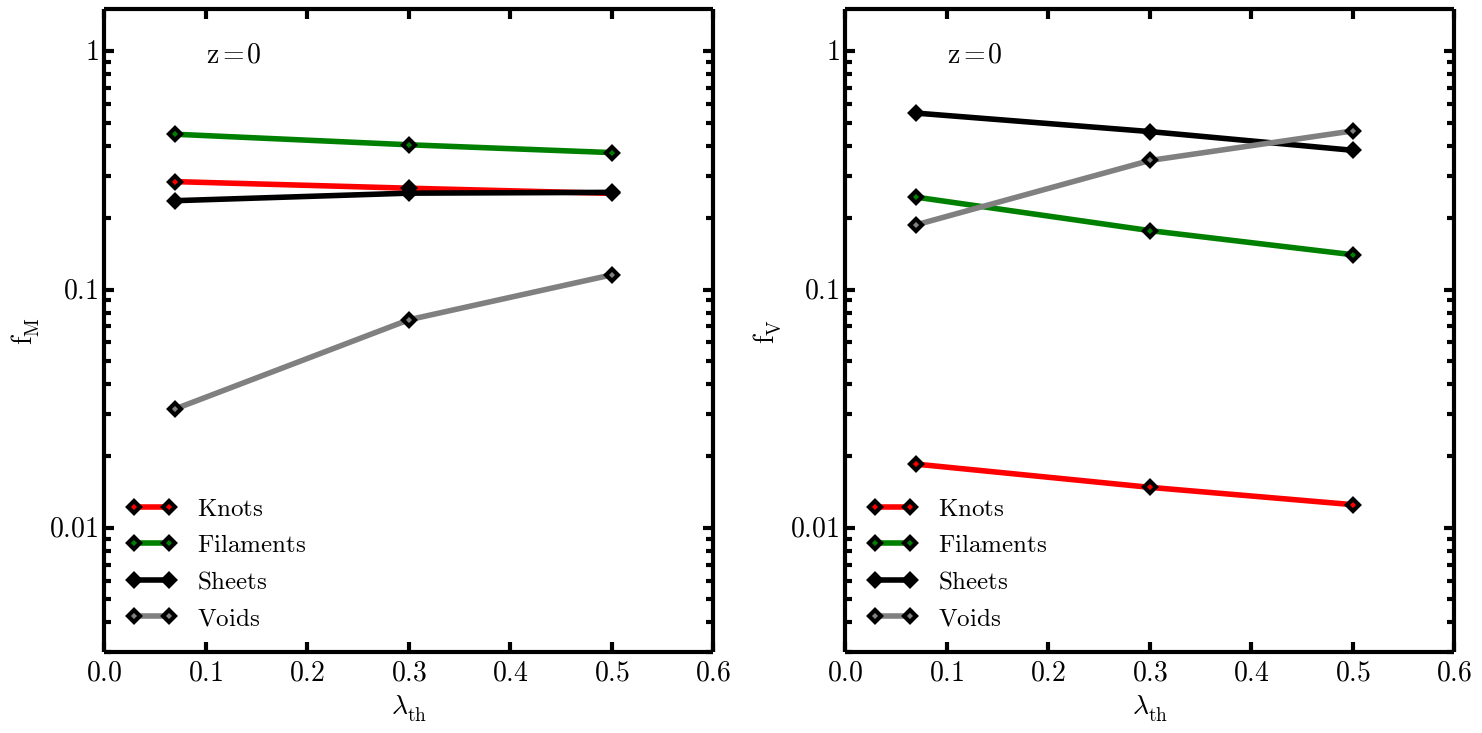}
\end{center}
\caption{Mass (left panel) and volume (right panel) fractions of voids, sheets, filaments and knots for different values of the eigenvalue threshold $\lambda_{\rm th}$ for the classification algorithm. Cosmic Web classification has been performed with smoothing scale $R_{\rm G}=4 \, {\rm cMpc}/h$ and grid size $512^3$ for this plot.}\label{fig:conv_lth}
\end{figure*}

In the first test, we kept the Gaussian smoothing scale fixed to 4 cMpc/$h$ and varied the resolution of the Cartesian grid used for the classification between $256^3$, $512^3$ and $1024^3$. Figure~\ref{fig:conv_resolution} shows how the mass and volume fractions of each class of structures vary with grid size. Most of the variation is seen passing from $256^3$ to $512^3$, whereas the results converge for higher grid size. Convergence is achieved when the Gaussian smoothing scale is at least 10 times larger than the Cartesian cell size. We concluded that our classification scheme achieves $\sim 1\%$ resolution at the fiducial $512^3$ resolution.

In the second test, we kept the Cartesian grid size fixed to $512^3$ and varied the Gaussian smoothing scale adopted for the density field between ${\rm R_G = 2}$ and 4 cMpc/$h$. The results of this test are reported in Figure~\ref{fig:conv_smoothing}. We only found relatively small variation in the volume fractions of voids and knots that are more sensitive to smoothing, given their usually sharp boundaries in 3-d space. The mass and volume fractions of different classes of structures did not vary significantly when we changed the smoothing length from 2 to 4 cMpc/$h$, therefore, we took ${\rm R_{G}=4}$ cMpc/$h$ as our fiducial choice. 

As a final test of robustness of the classification algorithm, we varied the value of $\lambda_{\rm th}$ at fixed Cartesian grid size ($512^3$) and smoothing length (${\rm R_{G}=4}$ cMpc/$h$). The $\lambda_{\rm th}$ threshold is the parameter that regulates the size of the eigenvalues of the deformation tensor that characterises collapse along a given direction. For this reason, changing this parameter should have significant influence on separating between voids, regions that did not collapse along any axes, and the other structures, regions that collapsed along at least one axis. As shown in Figure~\ref{fig:conv_lth} this is exactly what happens in our tests: the mass and volume fraction of knots, filaments and sheets do not vary significantly for $0.07\leq \lambda_{\rm th}\leq 0.5$. On the other hand, the value of $\lambda_{\rm th}$ has significant influence on the mass and volume fraction of voids. \cite{2009MNRAS.396.1815F} suggest $0.2\leq \lambda_{\rm th}\leq 0.4$ so that the volume fraction of voids is $>0.4$. For this reason we choose a fiducial value $\lambda_{\rm th}=0.3$. Since voids are expected to contain the lowest amount of mass, we do not expect our results on the baryonic gas budget in different phases to be strongly influenced by the value of $\lambda_{\rm th}$.

%%%

\section{Mass Fractions of Cosmic Structures in TNG100 vs. TNG300}\label{appendix:boxsize}

\begin{figure}
\begin{center}
 \includegraphics[width=0.48\textwidth]{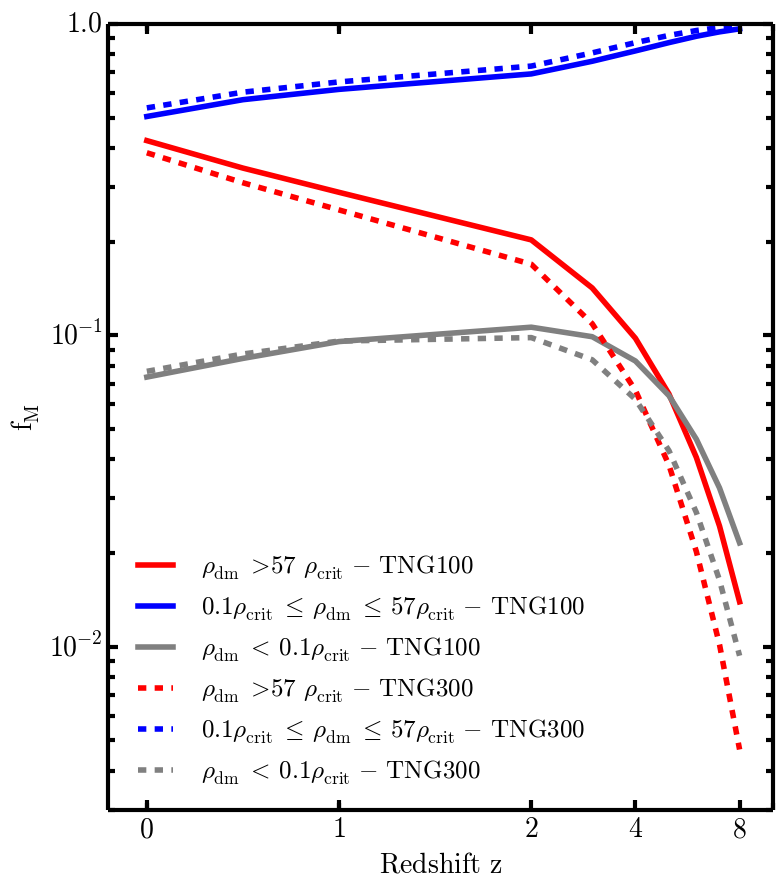}
\end{center}
\caption{ Comparison of the mass fractions associated to regions with different local dark matter density in TNG100 (solid lines) vs. TNG300 (dashed lines). Cosmic Web classification has been performed using the local dark matter density method. }\label{fig:100vs300}
\end{figure}

We compared the TNG100 and TNG300 simulations which use the same model for galaxy formation and feedback, but have different resolution and box size. The box volume of TNG300 is $\sim 20$ times larger than that of TNG100. The spatial and mass resolution of TNG100 are a factor 2 and 8 better than TNG300, respectively. In principle, the results of the Cosmic Web classification of the two simulations could differ due to (I) the different resolution which influences sampling small density fluctuations during the early evolution of the universe, and (II) the different box size which influence how many structures of each type are sampled. Comparison of the mass fraction associated to each cosmic structure in the two runs allows us to asses the robustness of our fiducial results from TNG100. Here we use a method based on the kernel-smoothed local dark matter density described in Subsection 3.1, This method is slightly less sophisticated than our fiducial classification method based on the deformation tensor, but is accurate enough to quantify the effect of box size on the properties of the Cosmic Web. The result of the comparison is shown in Figure~\ref{fig:100vs300}. We found that the mass fractions at redshift $z<4$ do not differ significantly between the two runs. We concluded that the volume of TNG100 is sufficiently large to be used as the fiducial simulation to perform our study.

\end{document}